\documentclass[useAMS,usenatbib]{mn2e}
\usepackage{graphicx}

\def\la{\raise.5ex\hbox{$<$}\kern-.8em\lower 1mm\hbox{$\sim$}}
\def\ma{\raise.5ex\hbox{$>$}\kern-.8em\lower 1mm\hbox{$\sim$}}

\def\Lsol{L$_{\odot}$ }

\def\kms{$\rm km\, s^{-1}$}
\def\cm3{$\rm cm^{-3}$}
\def\Ts{$\rm T_{*}$~}
\def\Vs{$\rm V_{s}$~}
\def\n0{$\rm n_{0}$}
\def\B0{$\rm B_{0}$}

\def\Te{$\rm T_{e}$}

\def\erg{$\rm erg\, cm^{-2}\, s^{-1}$}
\def\mum{$\mu$m~}

\def\L12{L$_{12\mu m}$~}
\def\F12{F$_{12\mu m}$~}

\def\Hb{H${\beta}$~}
\def\Ha{H${\alpha}$}

\def\RO3{R$_{[OIII]}$}

\title[LGRB host galaxies at  z$<$0.5]{
Metallicities in long GRB host galaxies at z$<$0.5
calculated by the detailed modelling of optical and infrared line ratios 
} 

\author[M. Contini]{M. Contini 
\\
School of Physics and Astronomy, Tel Aviv University, Tel Aviv
69978, Israel \\
}

\begin{document}


\pagerange{\pageref{firstpage}--\pageref{lastpage}} \pubyear{2009}

\maketitle

\label{firstpage}

\begin{abstract}

We revisited the line spectra emitted from long GRB (LGRB) host galaxies at z $\leq$ 0.5
in order to calculate by  the  detailed modelling of the line ratios, the
 physical conditions and relative abundances in LGRB hosts
 in this redshift range.
 We have found  lower metallicities than in LGRB hosts at higher z.
New results about metallicities and  physical conditions in the different regions  throughout the LGRB
980425 host at z=0.0085 are presented.
In particular, we have found that  the effective starburst temperature in the supernova (SN) region is the highest
throughout the host galaxy.
The low  ionization parameter reveals that  the radiation source
is far or somehow prevented  from reaching the emitting gas in the SN region.
The models constrained by a few oxygen, nitrogen  and sulphur  line ratios to \Hb in LGRB 980425
host satisfactorily reproduce   the  HeII/\Hb and [ArIII]/\Hb line ratios.
The  modelling of  the observed  [SIV]10.51\mum/[SIII]18.71\mum and [NeIII]10.6\mum/[NeII]12.81\mum line ratios  
from LGRB 031203 host galaxy at z=0.105  shows that the mid-IR lines are emitted from geometrically 
thin shock dominated filaments
which are not reached by the photoionizing flux, while the optical lines are emitted from the
radiation dominated outflowing clouds.

\end{abstract}

\begin{keywords}
radiation mechanisms: general --- shock waves --- ISM: abundances --- galaxies:  GRB  --- galaxies: high redshift

\end{keywords}

\section{Introduction}

Long duration $\gamma$-ray bursts (LGRB) 
derive  from  the death of very massive star 
(e.g. Paczynski 1998), they are flashes of cosmic 
high energy ($\sim $1  ${\rm keV} - 10$ GeV) photons (Fishman \& Meegan 1995) and
 explode in star forming galaxies. 
 LGRB and their afterglows
 are  associated with broad lined SN Ic (e.g. Hjorth et al 2003, Stanek et al 2003).
The analysis of LGRB host galaxy emission lines provides  information about star-forming galaxies at high z
(e.g. Kr\"{u}hler et al 2015,  Blanchard et al 2015 and references therein).
The fluxes of significant oxygen lines, a few nitrogen lines as well as \Ha~ and \Hb    are available 
from the surveys  at 
redshifts z$\leq$3. A few lines  appear in the literature at higher z.
 The strongest lines in the optical range  are [OII]3727+3729 (hereafter
[OII]3727+), [OIII]5007+ 4959 (hereafter [OIII]5007+) and [NII]6584   together with  \Hb and \Ha.
The [OIII]4363 line, which plays a dominant role in the modelling process,   is weak and not always available.
Some spectra contain also  [NeIII]3869,  [SII]6717, [SII]6731  and  seldom He, Fe and Ar lines
(e.g. Hammer et al 2006).  
The interpretation of the spectra leads to more or less converging theories about 
the distribution on z of star formation rates, ages, star masses and temperatures. The most interesting
parameters are related to metallicities,  such as the O/H relative abundances, followed by the
N/H one. 
Metallicity is one of the  main parameters  which affects the evolution of massive stars as
well as their explosive deaths  (Vergani et al 2011, Piranomonte et al 2015, Sollerman et al 2005, Woosley 1993, etc).
Niino et al (2016) claim that the relation between metallicity and  LGRB occurrence rate is not
understood quantitatively because, even when the redshifts of the host galaxies are well known,
the host galaxy is "not studied in detail".
Such a crucial issue, which yields many  consequences  in a large  z range deserves a more precise analysis.

In a previous paper (Contini 2016) we  presented new results  calculated by the detailed modelling of LGRB line ratios
 and we compared  them  with those  obtained by different investigations, 
in particular  for the element abundances in GRB host galaxies
at relatively high redshifts.
  The  emission line   spectra emitted from   LGRB host galaxies  
 on a large z range were   collected from the  surveys
 of Kr\"{u}hler et al,  Savaglio et al (2009),  Sollerman et al (2005), Castro-Tirado et al (2001), 
Graham \& Fruchter (2013), Levesque et al (2010), Vergani et al (2011), Piranomonte et al (2015), 
the LGRB line and continuum  spectra with Wolf-Rayet (WR) features from the Han et al (2010) survey.
In particular, the GRB 980425 host spectrum (Sollerman et al 2005)   was selected  because the survey  refers to 
relatively low redshift galaxies (z$<$ 0.2) that host GRB.
Sollerman et al  claim that they  are star-forming galaxies  with luminosities L$<$ \Lsol and   relatively 
low metallicity. 
 The spectra include the He I 5876 line which is also significant  in SN Type 1c hosts relatively to WR stars.

Contini (2016, fig. 6  top diagram)  shows the results of O/H and N/H calculated in each of 
the LGRB hosts by the detailed modelling of the spectra.
O/H  is close to solar in most objects. We will  conventionally  define   "solar"  relative abundances 
 (O/H)$_{\odot }$=6.6 - 6.7 $\times$10$^{-4}$ and (N/H)$_{\odot }$= 9.$\times$ 10$^{-5}$ 
 (Allen 1976, Grevesse \& Sauval 1998) that were found  suitable  to local galaxy nebulae. Moreover, these values 
 are included  between  those of Anders \& Grevesse (1989) 
(8.5$\times$10$^{-4}$ and 1.12$\times$10$^{-4}$,respectively)   and   Asplund et al (2009)
(4.9$\times$10$^{-4}$ and 6.76$\times$10$^{-5}$, respectively ).
Regarding nitrogen, we note a large distribution of N/H abundance ratios from solar to lower than solar    by a 
factor $>$10   in GRB hosts.
Subsolar N/H  indicates external gas acquisition through galaxy merging processes.
 Some observed LGRB hosts, however, show high metallicity. Kr\"{u}ler et al (and references therein)  point out that
"several metal rich GRB were discovered". Graham et al (2015) recently confirmed that high metallicities 
can be found in LGRB.
The near solar O/H relative abundances calculated from the observed line ratios by Contini (2016) in  
LGRB hosts exceed  the O/H  values  generally obtained by  other modelling methods. 
 In particular  at high z,  lower than solar O/H  in the host gas (e.g. Kr\"{u}hler et al 2015)
were   predicted by modelling the [OIII]/\Hb and [OII]/\Hb line ratios  using the  strong-line methods  (see Sect. 2).  
These results  are   generally obtained by the author majority.  
This issue (see Contini 2014 and references therein) will be  explained in the following section.

The  number of LGRB galaxy host spectra  at relatively low z investigated by Contini (2016)  was 
insufficiently small.
In this paper we try to  fill this gap, although by a relatively small number of objects,  by  modelling 
in detail  the LGRB host spectra  presented by Niino et al (2016) at z$\leq$ 0.41.
They  explain that the sample number  at these redshifts  cannot be large because the cosmic LGRB rate
density is low. On the other hand,  the faint galaxies  can be investigated
without very deep observations and the host galaxy spectroscopy can give information even when 
GRB afterglows are not bright enough.
In addition, low z GRB host galaxies give a hint about the  host characteristics in an epoch close to that of 
local galaxies,
connecting low  with high z objects.  We will investigate  this correlation.

 The spectral lines of some  galaxies in the Niino et al sample were reported by various  observers.
They may contain different information  in  each sample context, therefore
we refer to  each survey  as  presented by the observers.
For galaxy GRB 980425 host Niino et al refer to the average spectrum presented by Christensen et al (2008), who report
for this object at z=0.0085  a rich collection of spectra in different locations  within the host.
We have the rare occasion to calculate  the distribution of the physical
quantities throughout  a galaxy at  redshifts higher than local by  modelling each of the observed spectra.
Moreover, previous spectra presented by Hammer et al (2006)  for the GRB 980425 SN, WR star and  the  "4" regions  
show many lines which can  definitively constrain the models and confirm the presence of the WR stars.

The strongest lines  observed in  recent years  from  galaxies at relatively high z  are  in the optical range.
Observations in the infrared (IR)  are now available e.g. from Michalowski et al (2016) and Watson et al (2016)
showing  strong [CII]158\mum, [OI]63\mum in the far-IR (FIR) and [SIV]10.51\mum, [SIII]18.71\mum, 
[NeIII]15.56\mum, [NeII]12.81\mum, etc.  lines in the  mid-IR, respectively.
In this paper we will investigate whether the optical lines and the mid-IR and FIR lines  from the 
GRB 0301203 host galaxy are emitted  from  clouds of gas in similar physical conditions.

We use for the calculation of the line ratios the code SUMA which simulates  an emitting 
gaseous cloud within
a  galaxy hosting an AGN or a starburst, heated and ionized by the coupled effect of the primary and 
secondary photoionization fluxes and shocks.  
 Shocks throughout  the host galaxies  are the product of   star explosions and of
cloud collisions by e.g. galaxy merging (Contini 2016 and references therein). 
The "detailed modelling" method is  briefly explained  and compared with the    strong-line methods in Sect. 2.
The analysis of Niino et al observed line spectra is presented in Sect. 3 and  those of  Christensen et al
and Hammer et al relatively to the LGRB 980425 host galaxy are investigated in Sect. 4.  The optical and IR emission
lines observed from LGRB 031203 host  are discussed in Sect. 5.
 Concluding  remarks follow in Sect. 6.

\begin{table*}
\centering
\caption{Modelling [OII]3727+, [OIII]5007+, \Ha~ and [NII]6583  line ratios to \Hb from  Niino et al (2016) LGRB host galaxy spectra}
\begin{tabular}{lcccccccccccccccc} \hline  \hline
\           &z     & [OII]/& [OIII]/& \Ha/ & [NII]/ &\Vs  &\n0  & $D$        & O/H     &N/H      & \Ts   &$U$  & \Hb \\
\           &      &\Hb   & \Hb   &\Hb  & \Hb  &\kms &\cm3 &10$^{18}$cm &10$^{-4}$&10$^{-4}$&10$^4$K & -   & $^1$\\ \hline\ 980425$^2$&0.0085&5.39  & 3.86  &4.02 & 0.44  &-    &-    &-           &-        &-        &-      &-    &-    \\
\ mN1a      &   -  &5.65  & 3.86  &2.97 &0.42   &150  & 80  &0.4         &5.       &0.17     &9.     &0.0066&0.0041\\
\ 980425$_c^3$&0.0085&7.09  & 3.73  &3.00 & 0.32  &-    &-    &-           &-        &-        &-      &-    &-    \\
\ mN1b      &   -  &7.03  & 3.93  &2.99 &0.34   &150  & 80  &0.4         &5.3      &0.12     &10.    &0.0052&0.0034\\
\ 060505$^4$&0.089 &3.93  & 1.72  &5.13 &1.15   &-    &-    &-           &-        &-        &-      &-     &-     \\
\ mN2a      & -    &4.1   & 1.77  &2.95 &1.2    &200  &170  &0.1         &4.9      &0.3      &6.5    &0.009 &0.012 \\
\ 060505$_c$&0.089 &6.51  & 1.62  &3.00 &0.676  &-    &-    &-           &-        &-        &-      &-     &-     \\
\ mN2b      & -    &6.3   & 1.64  &2.99 &0.6    &200  &170  &0.1         &6.4      &0.2      &7.0    &0.005 &0.07  \\
\ 031203$^{5,6}$&0.105 &1.06& 8.48&2.82 &0.15   &-    &-    &-           &-        &-        &-      &-     &-     \\
\ mN3       &  -   &1.5   &8.4    &2.9  &0.17   &170  &250  &1.3         &4.5      &0.15     &7.7    &0.88  &0.19  \\
\ 060614$^7$& 0.
125&4.15  &2.2    &3.1  &$<$0.2 &-    &-    &-           &-        &-        &-      &-     &-     \\
\ mN4       & -    &4.4   &2.2    &2.95 &0.24   & 190 &170  &0.12        &4.7      &0.1      &7.     &0.009 &0.012 \\
\ 030329$^8$&0.169 &1.37  &5.     &3.19 &0.06   &-    &-    &-           &-        &-        &-      &-     &-     \\
\ mN5       &-     &1.45  &5.     &2.94 &0.06   &120  & 90  &0.5         &5.       &0.1      &6.3    &0.09  &0.018  \\
\ 120422$^9$&0.238 &4.53  &2.6    &4.18 &0.6    &-    &-    &-           &-        &-        &-      &-     &-      \\
\ mN6a      &-     &4.6   &2.65   &2.95 &0.6    &190  &170  &0.12        &4.8      &0.25     &7.5    &0.0095&0.012   \\
\ 120422$_c$&0.238 &6.19  &2.5    &3.00 &0.43   &-    &-    &-           &-        &-        &-      &-     &-      \\
\ mN6b      &-     &6.2   &2.53   &2.96 &0.42   &190  &170  &0.12        &6.1      &0.15     &7.7    &0.0069&0.0093  \\
\ 050826$^{10}$&0.296 &2.89  &1.73   &3.09 &0.52   &-    &-    &-           &-        &-        &-      &-     &-      \\
\ mN7       &-     &2.9   &1.8    &2.94 &0.58   &160  &250  &0.3         &4.9      &0.3      &5.5    &0.015 &0.026  \\
\ 130427A$^{11}$&0.340&2.91  &2.47   &4.15 &0.62   &-    &-    &-           &-        &-        &-      &-     &-   \\
\ mN8a      &-     &2.9   &2.44   &2.94 &0.58   &160  &250  &0.3         &5.6      &0.4      &5.5    &0.02  &0.033  \\
\ 130427A$_c$&0.340&3.94  &2.38   &3.00 &0.44   &-    &-    &-           &-        &-        &-      &-     &-      \\
\ mN8b      &-     &3.93  &2.3    &2.94 &0.6    &160  &250  &0.3         &6.2      &0.3      &6.0    &0.012 &0.022  \\
\ 130427A$^{12}$&-    &4.2  &1.8    &3.   &0.3    &-    &-    &-           &-        &-        &-      &-     &-      \\
\ mN9       &-     &4.1   &1.8    &2.96 &0.46   &160  &250  &0.36        &5.8      &0.2      &6.6    &0.008 &0.016  \\
\ 061021$^{13}$&0.346 &3.2&4.2    &3.8  &$<$0.52&-    &-    &-           &-        &-        &-      &-     &-      \\
\ mN10      &  -   &3.4   &4.1    &3.   & 0.54  &120  &280  &0.6         &5.8      &0.2      &7.6    &0.016 &0.034  \\
\ 011121$^{14,15}$&0.362 &3.38&1.95  &3.83 &0.17   &-    &-    &-           &-        &-        &-      &-     &-      \\
\ mN11      &-     &3.2   &1.98   &2.94 &0.19   &160  &250  &0.3         &5.3      &0.1      &5.5    &0.015 &0.026  \\
\ 1207149$^{16}$&0.398 &3.92&4.24 &3.   &0.24   &-    &-    &-           &-        &-        &-      &-     &-      \\
\ mN12      &-     &3.73  &4.25   &3.   &0.26   &120  &280  &0.6         &6.2      &0.14     &7.6    &0.016 &0.035  \\ \hline
\end{tabular}

$^1$ in \erg  (calculated at the nebula);
$^2$ Christensen et al (2008);
$^3$  the subscript c refers to the corrected line ratios;
$^4$ Th\"{o}ne et al (2008);
$^5$ Prochaska et al (2004);
$^6$ Sollerman et al (2005);
$^7$ Niino et al (2016);
$^8$ Levesque et al (2010a);
$^9$ Schulze et al (2014);
$^{10}$ Levesque et al (2010b);
$^{11}$ Niino et al (2016);
$^{12}$ Kr\"{u}ler et al (2015);
$^{13}$ Kr\"{u}ler et al (2015);
$^{14}$ Garnavich et al (2003);
$^{15}$ Graham \& Fruchter (2013);
$^{16}$ Kr\"{u}hler et al (2015);
\end{table*}

\section{Detailed modelling of the line spectra}

\subsection{Comparison of modelling methods}

To calculate  the O/H abundance ratio from the observations,
 "direct methods" (see e.g. Modjaz et al 2008)  are generally adopted.
They date back to the early  '70ties when  active galaxy line spectra
started to appear  (see Seaton 1975, Pagel et al 1992). Later, they were  updated.
 Determination of the O/H relative abundance from the oxygen line ratios to \Hb by 
current  strong-line methods
focus on  [OIII]/\Hb and [OII]/\Hb. 
 The  line fluxes depend mostly on the fractional
abundance of the ions and on the relative abundance of the elements. In particular, for intermediate
ionization- level lines such as [OII] and [OIII], the fractional abundance of the relative ions
are high in the radiation dominated zone of the cloud, while strong shocks are  necessary to obtain
strong lines from high ionization-levels (see, e.g. Sect. 5) and  recombined element lines.
By detailed modelling,  the line fluxes are integrated throughout the recombination region where
 a large zone of gas  is characterized by  \Te $<$10$^4$ K (see Sect. 5).  
O$^+$/O and O$^{2+}$/O fractional abundances peak at different temperatures. In the cool gas region
both  are relatively low. Therefore, to reproduce high
 [OII]/\Hb and [OIII]/\Hb line ratios, relatively high O/H are invoked.
By the   strong-line models, a single temperature  of 10$^4$K is generally adopted
corresponding to high O$^+$/O and O$^{2+}$/O fractional abundances, so
 relatively  low O/H   are suitable  to fit the observed line ratios.
It was explained by Contini (2014) that the metallicities in terms of  O/H and N/H
obtained by  the  strong-line methods are lower limits.
Moreover, less  consistent results are obtained   comparing the data with diagnostic diagrams 
calculated for general cases.
For  spectra rich in number of lines from
different elements in  various frequency ranges,
the CLOUDY, SUMA and other  codes  were assembled. SUMA accounts
 for both the photoionization and the shock because 1) it was noticed that  lines from recombined elements and
from very
high ionization levels could not be  reproduced consistently using   photoionization alone,
2)  the radio continuum shows the synchrotron  power-law
created by the Fermi mechanism at the shock front in most of   the spectral energy distribution (SED),
3)  low [OIII]5007+/[OIII] 4363 line ratios
(such as observed in LINERs) could not be reproduced without the shocks and
above all,  because 4) in galaxies at high z, that  derive mostly from mergers, shocks are created by
collision, etc.
By the detailed modelling method  the simulated emitting clouds in the galaxy
are characterized by the parameters of the shock and of the
photoionization, by the element abundances , by the geometrical thickness due to fragmentation, etc.
We calculate the emission lines from  the gas  adopting a set of the most
significant input parameters.  The calculation process is  briefly described in the following.
 Some of the parameters  are suggested by the observations (e.g. the FWHM gives a hint to
the shock velocity choice, the electron density is roughly deduced from the [SII] 6716/6731 doublet ratio, etc)).
Moreover, some published grids of model results (Contini \& Viegas 2001a,b)  indicate in a general way how to 
choose the first set of input parameters.
The lines calculated by the code are more than 200, 
 because  line emission contributes to the cooling process of the emitting gas  in the recombination zone.
The number of calculated lines is independent from the observations.
Indeed, when only a few lines are observed the  initial parameter set is less constrained.
So we create a grid of models in order to  reproduce the observations
and to avoid degeneracy.
In the modelling process, we   aim to reproduce the  observed line ratios  for each element.
Each line has a different  strength which  translates into  the different precision  by the fitting process.
A minimum number of significant lines ([OIII] 5007+,  [OII]3727+, [OIII]4363, [NII], \Ha, \Hb) is
necessary to constrain the model.
We deal with  line ratios to avoid  distance and morphological effects.
A perfect fit of the observed line ratios is not realistic because
the observed data have errors, both random and systematic.
The set of parameters which leads to the best fit of the observed line ratios
and continuum SED, is  regarded as the result of modelling.
 The results are acceptable when the observed  strongest  line ratios are reproduced  within  10\%, 
and the weakest by $\sim$ 50 \%.

\subsection{Brief description of the  calculations}

By the  SUMA code (Contini 2015 and references therein) 
 line and continuum emissions
from the gas are calculated consistently with dust-reprocessed radiation in a plane-parallel geometry.
 The calculations start at the shock front where the gas is compressed and thermalized adiabatically,
reaching the maximum temperature in the immediate post-shock region
($T(K)\sim 1.5\times 10^5 (V_{\rm s}/100$ km s$^{-1}$$)^{2}$,
where \Vs is the shock velocity).
T decreases  downstream following the cooling rate.
The input parameters such as  \Vs, the atomic preshock density \n0 and
the preshock magnetic field \B0 (for  all models \B0=10$^{-4}$Gauss is adopted)
define the hydrodynamical field.  They  are  used in the calculations
of the Rankine-Hugoniot equations  at the shock front and downstream. They  are combined in the
compression equation which is resolved  throughout each slab of gas in order to obtain the density
profile downstream.
 The input parameters  that represent the primary radiation
from  the host starburst (SB) are  the effective temperature \Ts
 and the ionization parameter $U$.
 A  pure black-body radiation referring to  \Ts  is a poor approximation  for a starburst,
even adopting  a dominant spectral type (see Rigby \& Rieke 2004). However, it is the most suitable because the
line ratios  that  are used to indicate \Ts also depend on
metallicity, electron temperature, density, ionization parameter, the
morphology of the ionized clouds,  and, in particular, they depend on the hydrodynamical field.
  The primary radiation source
is independent  but it affects the surrounding gas.
In contrast, the secondary diffuse radiation is emitted from the slabs of gas heated
by the radiation flux reaching the gas and collisionally  by the shock.
In our model the gas region surrounding the radiation source is not considered as a
unique cloud, but as an ensemble of fragmented filaments. The geometrical
thickness of these filaments is  an input parameter of the code ($D$) which is
calculated consistently with the physical conditions and element abundances of
the emitting gas.
Primary  and  secondary radiations are  calculated by radiation
transfer throughout the slabs downstream.
The fractional abundances of the ions are calculated resolving the ionization equations
for each element in each ionization level.
The dust-to-gas ratio ($d/g$) and the  abundances of He, C, N, O, Ne, Mg, Si, S, A, Fe, relative to H,
are also accounted for.
 The uncertainty in the calculations
is due to  the atomic parameters (within 10 \%)  which are
often updated.

\begin{figure}
\centering
\includegraphics[width=8.0cm]{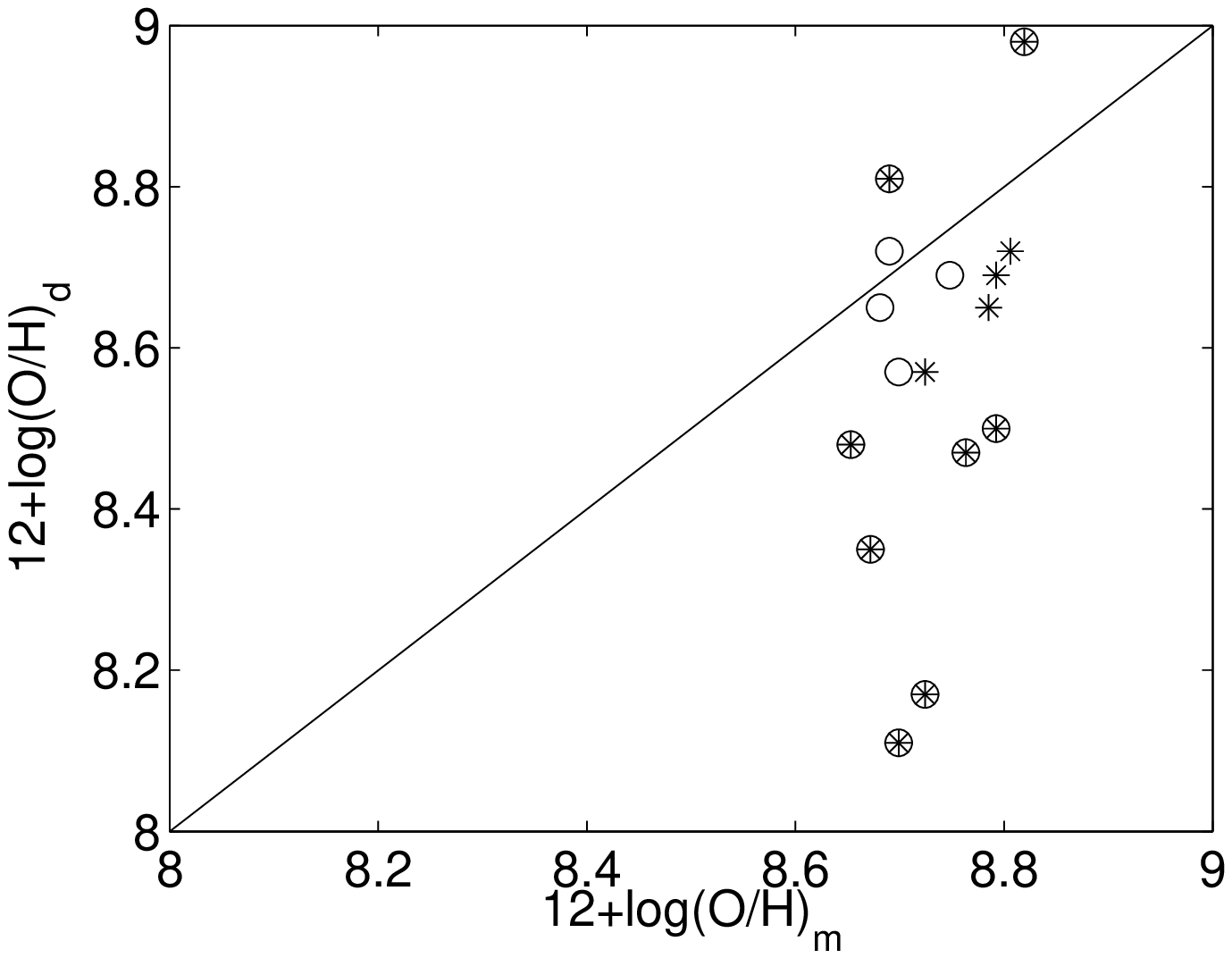}
\includegraphics[width=8.0cm]{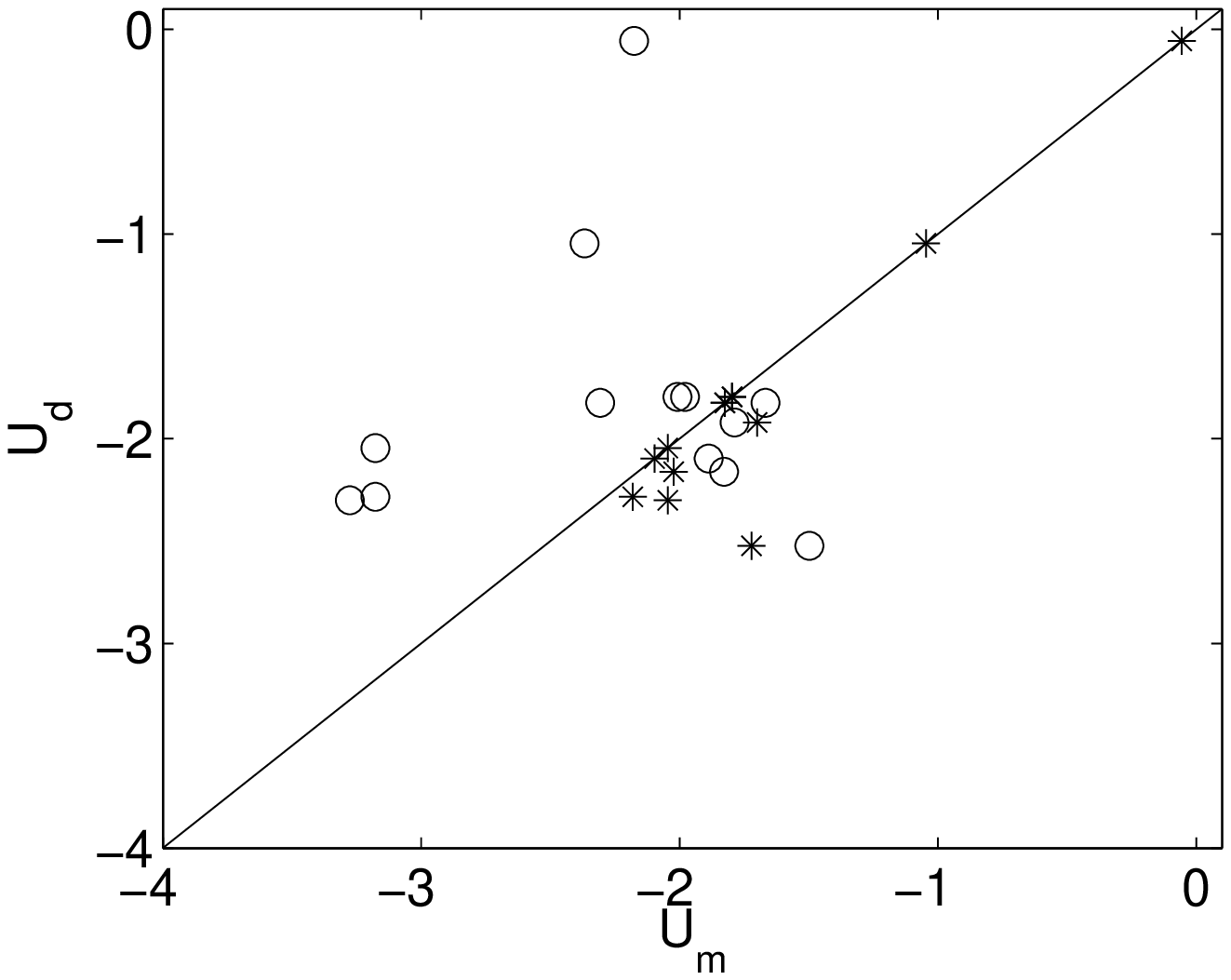}
\caption{Top : 
comparison of   12+log(O/H) calculated by detailed modelling (m)  with
those calculated  using 
indirect diagnostic calibrations (R23, N2)  by Niino et al (d).
Bottom : comparison of $U$  calculated by detailed modelling (m)  with
those calculated   by Niino et al (d).
open circles: model calculations; asterisks:
model calculations  referring to reddening corrected data
}
\end{figure}

\begin{figure}
\centering
\includegraphics[width=8.0cm]{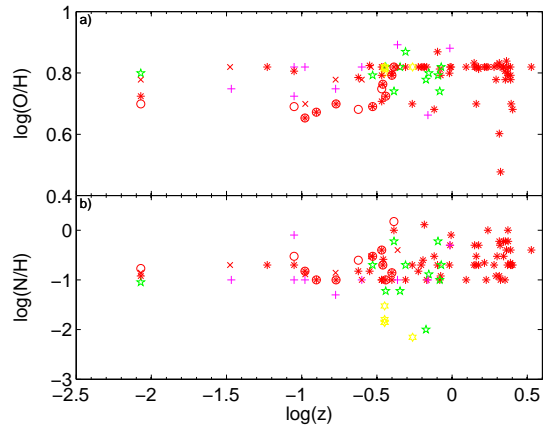}
\caption{Calculated log(O/H) (top) and log(N/H) (bottom)  (in units of 10$^{-4}$,
see  Table 1) as a function of z on an extended  z range :
symbols as in Fig. 1.  Moreover,
green stars :  LGRB  (Kr\"{u}hler et al);
yellow hexagrams : SGRB (de Ugarte Postigo et al);
magenta plus : LGRB with WR stars (Han et al).
}
\end{figure}

\section{Niino et al (2016)  sample galaxies}

Niino et al  presented spectroscopy results for three LGRB host galaxies at z$\leq$ 0.41, observed with 
Gemini Multi-Object Spectrograph (Hook et al (2004), together with data collected from the literature. 
 Some galaxy hosts were previously observed, but  
new data were presented by Niino et al.  because for
GRB 060614 host  the metallicity measurements were uncertain.
Therefore, it was revisited by GMOS-South spectroscopy. GRB 130427A host emission lines were already 
reported by Xu et al (2013) and Kr\"{u}hler et al (2015), but the [NII] detection  was marginal in both.
The host was revisited by Subaru/FOCAS.
GRB 111225A host has been analyzed by archival data (Th\"{o}ne \& de Ugarte Postigo 2014) but they were not reported.
The host was revisited with Subaru/FOCAS spectroscopy.
Recently,  Perley et al (2016) suggested that
  GRB 020819B host at z=0.410 reported by   Niino et al  
happens to lie close to the line of sight of GRB020819B at z=1.9621,
but  it  is entirely unrelated to the
GRB. So it was removed from  the modelling list.

In Table 1 we compare the  calculated with the  observed line ratios from Niino et al table 5.
The observed line spectra were corrected for  foreground Galaxy extinction.
The theoretical \Ha/\Hb line ratios at the emitting nebula should be $\sim$ 3, considering
that the emitting gas has a distribution of densities and temperatures  in the recombination zone downstream
of the shock front, and behind the ionization front  created by the  radiation source.
For some galaxies in the Niino et al sample the observed \Ha/\Hb are $\geq$ 4 , reaching
even values $\geq$ 5.15 and 5.5 for 060505 and
020819B, respectively. Such high \Ha/\Hb values  can be found in  high density gas ($>$ 10$^6$ \cm3) where
some self-absorption occurs in the Balmer lines (Osterbrock 1974). This leads to the strengthening of the \Ha~ line
relatively to the other lines of the Balmer series (see also Contini 2003).
 For the forbidden lines, in particular [OII], [SII] etc. the critical densities for  collisional deexcitation
 are $\geq 4\times10^3$ \cm3 and  for the [OIII] 5007+ lines $\sim$6.5$\times$10$^5$ \cm3,
therefore  high density gas  in the host galaxies could be revealed  by relatively strong permitted lines  
and   abnormally high \Ha/\Hb   ratios.
Alternatively, considering  that the
line fluxes are  affected by  gas and dust through their path to Earth, the data should by further
 reddening corrected.
In Table 1 first column the identification number of the galaxies is given  followed by the redshift and
by the  line ratios presented in  the different surveys, reported by Niino et al.
In the  rows next to  those referring to  observations, the best fitting calculated line ratios
and the parameters adopted in the selected models are  given in columns 3-6 and columns 7-13, respectively.
In the last column of Table 1 the \Hb line fluxes  calculated at the nebula are reported.
For  galaxies 980425 , 060505, 120422, 130427A and 020819B, we present both the data given by Niino et al
and those reddening corrected (e.g. 980425$_c$).
Models mN1, mN2, mN6, mN8 and mN13 split into e.g. mN1a and mN1b, where the former
refers to Niino et al data and the latter to the corrected line ratios.

The results of modelling  give shock velocities between 120 and 200 \kms, pre-shock densities
between 80 and 280 \cm3, in agreement with previous results obtained for a large sample of LGRB hosts
(Contini 2016). Except for 060614,  for all the galaxies the geometrical thickness of the emitting clouds
is $\leq$  0.2 pc.
The physical conditions of the gas emitting the Balmer lines  show ${\tau}$ $\leq$1.
  There is a rough agreement for $U$ calculated from modelling the reddening corrected spectra
and $U$=q/c  determined  on the basis of   strong-line methods  (Kobulnicky \& Kewley 2004) applied, however, to
undereddened line ratios by Niino et al (Fig. 1, bottom diagram).  
The  reddening corrected [OII]/[OIII] line ratios  are higher than those presented by Niino et al,
 implying  models with higher star temperatures, higher O/H and lower $U$, because these parameters
affect differently the  [OII]/\Hb and [OIII]/\Hb line ratios.
The O/H metallicities  calculated by the detailed modelling
and by the  strong-line method are compared in Fig. 1 (top diagram).
 The uncertainty in the Niino et al metallicity is $\sim$ 0.1.
  A rough agreement is seen  for 12+log(O/H)$>$ 8.65 
 considering the  approximation for both the   strong-line methods and the detailed modelling.

Even if higher O/H  are  predicted by the   detailed modelling than by  the  strong-line  methods,
with O/H  between  Allen solar value (6.6$\times$10$^{-4}$ ) and  the Asplund et al.  one (4.9$\times$10$^{-4}$),
 the O/H relative abundances calculated to fit the Niino et al corrected line ratios
are all  lower than solar. This would
imply that O/H ratios calculated by detailed modelling for the LGRB host sample  at z$<$0.41  are lower than
those (mostly solar)  calculated by Contini (2016)  for higher z objects   by a factor $\leq$2 , as shown in Fig. 2.
A few  LGRB hosts  with O/H  $\leq$ 4.9$\times$10$^{-4}$
 could be seen as  minima throughout the z range. The O/H ratios recover  solar values towards
local galaxies.
 Considering that  lower than solar metallicities derive from  mixing with
external matter during galaxy merging,  it seems that the merging process is highly efficient
at low z.

\section{The  spectra  in the different regions of  LGRB 980425 host galaxy}

Niino et al  report the average spectrum of the LGRB 980425 host galaxy presented by
 Christensen et al in table 2, last row.
The  uncertainties are very large in the line flux means, therefore in the following we   
refer to the single region spectra.

\subsection{Modelling  Christensen et al (2008)  data}

In Tables 2 and 3 we present the modelling of the spectra observed throughout the 980425 host
and the selected models  which best reproduce the observed line ratios.
The data were obtained by the Very Large Telescope UT3 {Melipal} with the VIMOS integral field
spectroscopy mode.
In Table 2 the line ratios  reported by Christensen et al are shown in columns 2-7
 followed by the reddening corrected line ratios in columns 9-14.
In the next rows the  calculated line ratios   best fitting the  data  (models mc1-mc26)
and best fitting the corrected line ratios (models m1c-m26c) are reported.
The  models   are described in Table 3.
Most of the observed \Ha/\Hb  are $>$3.
As discussed in Sect. 3, the pre-shock densities and the geometrical thickness used
by models mc1-mc26 yield optical thickness $\tau$$<$1, therefore we have  corrected  for reddening
the line ratios.
In Fig. 3 (top diagram)  the observed \Ha/\Hb are given  as function of the observed [OII]/\Hb.
The three observed spectra  corresponding to models mc6, mc11 and mc17 show  \Ha/\Hb = 12.7, 10.4 and 11.9,
respectively.
The correction factor is very high and leads to  high corrected  [OII]/\Hb, in particular for the latter,
which corresponds
to an  already high uncorrected [OII]/\Hb. High [OII]/\Hb are generally  found in  collisionally dominated
nebulae,   suggesting that shocks are at work.
The \Ha/\Hb ratios for the two former spectra (referring to mc6 and mc11) correspond to relatively
low [OII]/\Hb indicating that the reddening correction makes sense.
The [OIII]/\Hb versus [OII]/\Hb
corrected and not- corrected line ratios  (Fig. 3 bottom diagram) show that even if the trend corresponds
to a radiative situation with small $U$ (Contini 2016, fig. 1),
the data are scattered,  confirming that another mechanism, e.g. the shocks cannot be neglected.

\begin{figure}
\centering
\includegraphics[width=8.0cm]{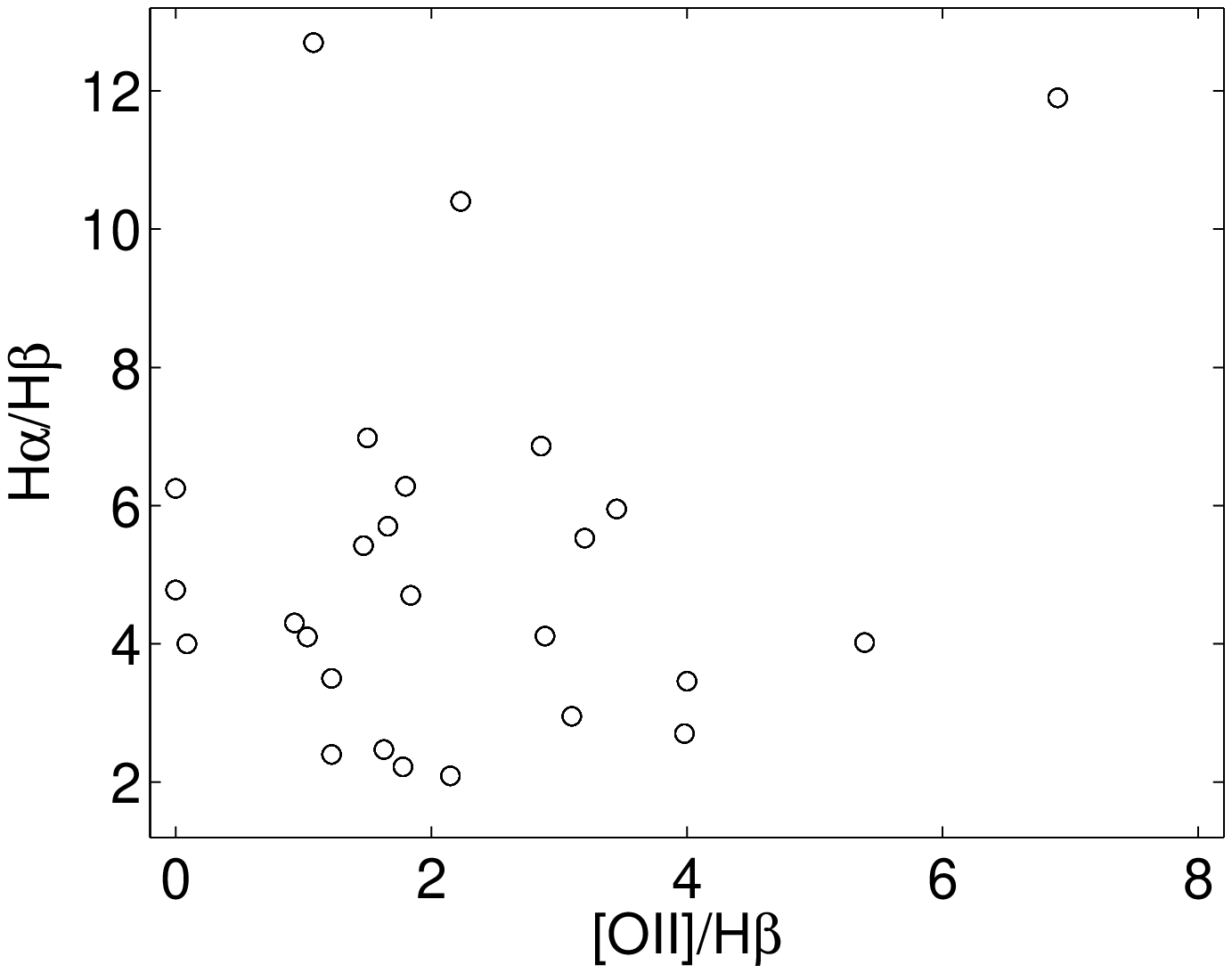}
\includegraphics[width=8.0cm]{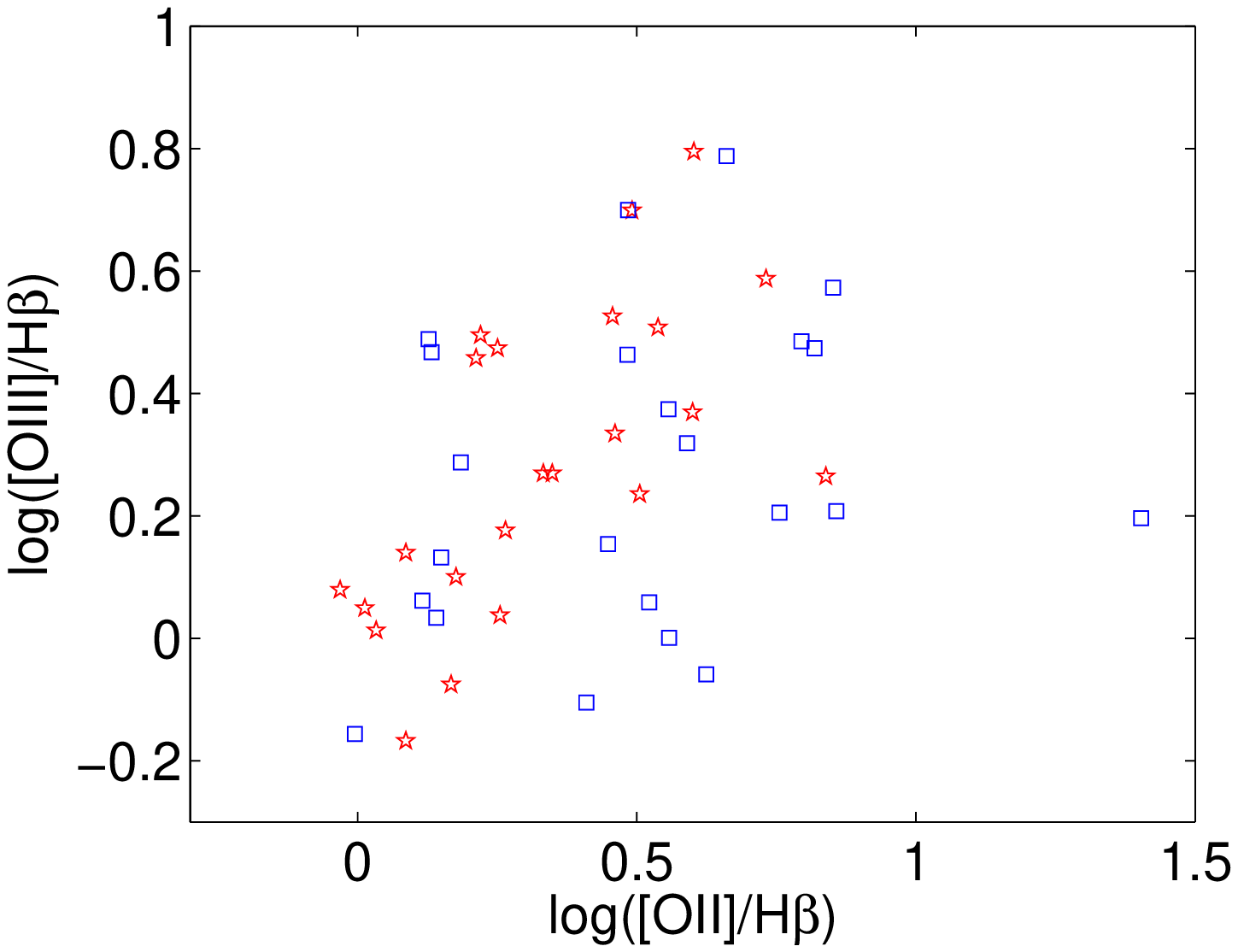}
\caption{Christensen et al data. Top : observed \Ha/\Hb versus observed [OII]/\Hb.
Bottom : observed  [OIII]/\Hb versus [OII]\Hb (red pentagrams);
corrected [OIII]/\Hb versus [OII]/\Hb (blue squares)}
\end{figure}

\begin{table*}
\centering
\caption{Modelling [OII]3727+, [OIII]5007+, \Ha, [NII]6583, [SII]6717, 6731 line ratios to \Hb=1 from  
  GRB 980425 host regions (Christensen et al 2008)}
\begin{tabular}{lccccccccccccc} \hline  \hline
\            & [OII]/& [OIII]/ & \Ha/ & [NII]/ & [SII]/ & [SII]/  & & [OII]/    & [OIII]/  & \Ha/ & [NII]/&[SII]/&[SII]/\\
\            &\Hb   & \Hb    &\Hb  & \Hb   & \Hb   & \Hb   &  &\Hb       & \Hb     &\Hb  & \Hb  &\Hb  &\Hb \\\hline
\  galaxy    &5.39  &3.87   &4.02 & 0.44  & 0.63  & 0.15 &   & 7.1      &3.74     &3.   & 0.33 & 0.45 & 0.11\\
\  mc1       &5.41  &3.97   &2.97 & 0.44  &0.58   &0.66  &m1c&7.2       &3.8      & 3.  &0.39  &0.45  & 0.5  \\
\ WR region  &4.    &6.24   &3.46 & 0.27  &0.3    &0.23  &   &4.57      &6.14     &3.   & 0.23 & 0.25 &0.2\\
\  mc2       &4.    &6.24   &2.94 & 0.3   &0.27   &0.3   &m2c&4.5       &6.18     & 3.  & 0.3  & 0.26 &0.3 \\ 
\ SN region  &3.45  &3.22   &5.95 &0.67   &1.22   &0.91  &   &6.6       &3.       &3.   & 0.34 & 0.56 & 0.4\\
\  mc3       &3.4   & 3.2   & 2.96&0.68   &0.78   &0.9  &m3c &6.7       & 3.2     & 3.  &0.36  & 0.54 &0.6 \\  
\ WR(-8.7)   &3.1   &5.     &2.95 &0.21   &0.21   &0.17 &    &3.        & 5.      & 3.  & 0.21 & 0.21 &0.17 \\
\ mc4        &3.    &5.2    &2.95 &0.2    &0.27   &3.   &m4c &3.        & 5.2     & 3.  & 0.22 &0.27  & 0.32\\   
\ SN(-5.3)   &2.86  &3.36   &6.86 &0.82   &1.53   &1.12 &    &6.24      &3.       & 3.  & 0.36 &0.59  & 0.42\\
\ mc5        &2.79  &3.33   &2.96 &0.86   &0.83   &0.9 &m5c  &6.4       &3.1      &3.   &0.24  &0.53  &0.56 \\
\ (-11.4)    &1.08  &1.03   &12.7 &0.146  &2.     &1.52 &    &4.2       &0.87     &3.   &0.034 &0.38  &0.28\\
\ mc6        &1.1   &1.01   &2.99 &0.17   &1.3    &1.5 &m6c  &4.3       &0.86     &3.   &0.1   &0.35  &0.38\\
\ (-8.0)     &1.8   &1.09   &6.28 &1.08   &1.65   &1.29 &    &3.6       &1.       &3.   &0.52  &0.71  &0.54 \\
\ mc7        &1.8   &1.12   &3.   &0.9    &1.64   &1.7 &m7c  &3.86      &1.       &3.   &0.6   &0.7   &0.7  \\
\ (-6.7)     &1.5   &1.26   &6.98 &0.86   &1.22   &1.04 &    &3.3       &1.1      &3.   &0.37  & 0.46 &0.39 \\
\ mc8        &1.7   &1.25   &3.   &0.8    &1.23   &1.29 &m8c &3.4       &1.19     &3.   &0.4   &0.48  &0.49 \\
\ (-4.0)     &1.84  &1.5    &4.7  &0.59   &0.73   &0.53  &   &2.81      &1.42     &3.   &0.37  &0.44  &0.3 \\
\ mc9        &1.9   &1.5    &2.98 &0.6    &0.6    &0.62 &m9c & 2.97     &1.4      &2.98 &0.36  &0.45  &0.46\\
\ (-2.0)     &3.2   &1.72   &5.53 &0.78   &0.94   &0.77 &    &5.7       &1.6      &3.   &0.42  &0.47  &0.37\\
\ mc10       &3.    &1.77   &2.98 &0.7    &0.8    &0.73 &m10c&5.88      &1.6      &3.   &0.4   &0.47  &0.42\\   
\ (-1.3,-6.1) &2.23  &1.86   &10.4 &1.2    &1.84   &1.29 &   &7.2       &1.6      &3.   &0.35  &0.44  &0.3\\
\  mc11       &2.1   &1.92   &3.   &0.9    &1.42   &1.22 &m11c&7.4      &1.6      &3.1  &0.46  &0.42  &0.35\\
\ (-1.3,10.0) &1.47  &0.84   &5.42 &0.76   &1.06   &0.86 &   &2.56      &0.78     &3.   &0.42  &0.53  &0.43 \\
\ mc12        &1.5   &0.84   &3.   &0.7    &1.07   &0.93&m12c&2.57      &0.82     &3.   &0.43  &0.58  &0.5\\ 
\ (-0.7)     &2.89   &2.16   &4.11 &0.53   &0.43   &0.3 &    &3.89      &2.08     &3.   &0.38  &0.3   &0.2\\
\ mc13       &2.96   &2.02   & 3.  &0.66   &0.46   &0.39&m13c&3.7       &2.1      &3.   &0.4   &0.3    &0.26 \\
\ (-0.0)     &1.66   &3.13   &5.7  &0.92   &1.23   &0.92&    & 3.0      &2.9      &3.   & 0.48 &0.59  & 0.43\\
\ mc14       &1.67   & 3.1   &3.   &0.9    &1.04   &1.04&m14c&3.1       &2.97     &3.   &0.6   &0.59  &0.58 \\ 
\ (2)        &0.0    & 0.99  &6.25 &0.87   &1.5    &1.28&    &0.0       &0.91     &3.   &0.42  &0.65  &0.54\\
\ mc15       & 0.7   &1.0    &3.   &0.7    &0.65   &1.18&m15c&0.8       &0.96     &2.97 &0.48  &0.46  &0.85\\ 
\ (2.7)      & 1.03  &1.12   &4.1  &0.61   &0.66   &0.44&    &1.38      &1.08     &3.   &0.44  &0.46  &0.30\\ 
\ mc16       &1.02   &1.13   &3.05 &0.63   &0.47   &0.51&m16c&1.34      &1.18     &3.   &0.4   &0.36  &0.4 \\
\ (2.7,-4.0) &6.9    &1.84   &11.9 &0.44   &0.62   &0.4 &    &25.       &1.57     &3.   &0.44  &0.62  &0.41\\
\ mc17       &7.0    &1.84   &3.17 &1.8    &2.6    &2.2 &m17c&25.6      &1.4      &3.25 &0.42  &0.55  &0.56\\
\ (5.4)      &0.93   &1.2    &4.3  &0.63   &0.77   &0.56&    &1.3       &1.15     & 3.  &0.44  &0.51  &0.37\\
\ mc18       &0.91   &1.25   &3.05 &0.5    &0.74   &0.79&m18c&1.45      &1.1      &3.1  &0.46  &0.57  &0.57 \\
\ (6.1)      &1.22   &1.38   &3.5  &0.53   &0.76   &0.54&    &1.4       &1.35     &3.   &0.45  &0.64  &0.45\\
\ mc19       &1.1    &1.4    &3.05 &0.6    &0.77   &0.8 &m19c&1.3       &1.35     &3.   &0.48  &0.52  &0.47 \\  
\ (8.7)      &1.22   &0.68   &2.4  &0.31   &0.68   &0.42&    &0.99      &0.7      &3.   &0.39  &0.88  &0.54\\
\ mc20       &1.21   &0.63   &3.   &0.4    &0.7    &0.6 &m20c&1.04      &0.7      &3.   &0.3   &0.7   &0.58  \\
\ (9.4)      &1.63   &2.87   &2.47 &0.27   &0.4    &0.3 &    &1.36      &2.93     &3.   &0.33  &0.5   &0.38 \\
\ mc21       &1.75   &2.9    &2.99 &0.4    &0.44   &0.44&m21c&1.4       &2.85     &3.   &0.32  &0.4   &0.42\\ 
\ (10.1)     &0.09   &2.45   &4.   &0.49   &0.68   &0.48&    &0.12      &2.37     &3.   &0.37  &0.49  &0.34 \\
\ mc22       &0.12   &2.47   &3.   &0.1    &0.02   &0.03&m22c&0.23      & 2.2     &3.   &0.4   &0.4   &0.5 \\
\ (11.4)     &1.78   &2.98   &2.22 &0.37   &0.33   &0.25&    &1.3       &3.08     &3.   &0.5    &0.46  &0.35\\
\ mc23       &1.8    &2.97   &2.99 &0.38   &0.37   &0.37&m23c&1.37      &2.95     &3.   & 0.44  &0.37  &0.37\\
\ (12.1)     &0.00   &1.3    &4.78 &0.67   &0.9    &0.77&    &0.0       &1.23     &3.   &0.42   &0.53  &0.45\\
\ mc24       &0.7    & 1.2   &2.98 &0.7    &0.66   &1.22 &m24c &0.8     &1.2      & 2.98&0.48   &0.4   &0.7\\
\ (14.8,3.3) &3.98   & 2.34  &2.7  &0.29   &0.56   &0.39 &   &3.6       &2.36     &3.   &0.47   &0.66  &0.61\\
\ mc25       &3.99   &2.24   &2.95 &0.3    &0.38   &0.5&m25c &3.67      &2.3      &2.95 &0.4    &0.45  &0.6\\
\ (14.8,10.) &2.15   &1.86   &2.09 &0.33   &0.44   &0.4 &    &1.53      &1.94     &3.   &0.47   &0.66  &0.61  \\
\ mc26       &2.3    &1.94   &3.   &0.5    &0.45   &0.39&m26c&1.5       &2.1      &3.   &0.48   &0.62  &0.53\\ \hline

\end{tabular}
\end{table*}

\begin{table*}
\centering
\caption{Models calculated to reproduce the  spectra by Christensen et al (2008) reported in Table 2}
\begin{tabular}{lcccccccccccc} \hline  \hline
\      &\Vs  &\n0  & $D$        & O/H     &N/H      &S/H        & \Ts   &$U$  & \Hb \\ 
\      &\kms &\cm3 &10$^{18}$cm &10$^{-4}$&10$^{-4}$&10$^{-4}$  &10$^4$K & -   &  $^1$\\ \hline
\ mc1  & 150 & 100 & 0.4        &4.9      &0.17     &0.04       &9.     &0.0076&0.006\\
\ m1c  &150  & 100 & 0.4        &6.       &0.14     &0.03       &9.4    &0.0056&0.0047\\
\ mc2  &150  &100  &0.4         &5.       & 0.17    &0.02       &8.4    &0.018 &0.01  \\
\ m2c  &150  &100  &0.4         &5.5      &0.17     &0.02       &8.4    &0.016 &0.01  \\
\ mc3  &150  & 100 &0.3         &5.5      &0.5      &0.08       &6.5    &0.014 &0.01  \\
\ m3c  &150  &100  &0.3         &6.3      &0.15     &0.04       &8.3    &0.0055&0.0046\\
\ mc4  &150  &100  &0.3         &6.6      &0.2      &0.03       &6.5    &0.032 &0.015\\  
\ m4c  &150  &100  &0.3         &6.6      &0.2      &0.03       &6.5    &0.032 &0.015\\ 
\ mc5  &120  &100  &0.3         &6.6      &0.8      &0.1        &6.     &0.018 & 0.01\\
\ m5c  &120  &100  &0.3         &6.6      &1.       &0.04       &8.4    &0.0044&0.0037\\
\ mc6  &120  &100  &0.3         &6.       &0.3      &0.3        &3.4    &0.15  &0.027  \\
\ m6c  &120  &100  &0.3         &6.       &0.06     &0.04       &5.     &0.003 &0.003\\
\ mc7  &130  & 80  &0.3         &6.6      &1.       &0.3        &4.4    &0.02  &0.0084\\
\ m7c  &130  &80   &0.3         &6.4      &0.4      &0.08       &5.     &0.005 &0.0034\\
\ mc8  & 130 & 80  & 0.3        & 6.4     & 1.      &0.22       &4.4    &0.024 &0.0095 \\
\ m8c  &130  &80   &0.3         &6.4      &0.3      &0.06       &4.8    &0.008 &0.0048\\
\ mc9  & 130 & 80  & 0.3        &6.2      &0.8      &0.1        &4.4    &0.03  &0.01   \\
\ m9c  &130  &80   &0.3         &6.4      &0.3      &0.06       &4.7    &0.012 &0.0063\\
\ mc10 & 120 & 70  & 0.2        &6.6      &0.6      &0.1        &5.2    &0.01  &0.0037 \\
\ m10c &120  &60   &0.2         &6.6      &0.2      &0.04       &5.9    &0.003 &0.0015 \\
\ mc11 & 120 & 50  & 1.         &6.7      &1.       &0.2        &5.5    &0.015 &0.0042  \\
\ m11c &120  &50   &1.          &6.6      &0.2      &0.03       &7.8    &0.0014&8.3e-4 \\
\ mc12 & 120 & 50  & 2.         & 6.3     &1.       &0.2        &4.4    &0.015 & 0.0046 \\
\ m12c&120  &50   &2.          &6.2      &0.4      &0.08       &4.8    &0.0055&0.0026\\
\ mc13 & 120 & 50  &2.          &6.3      &0.6      &0.05       &6.1    &0.008 &0.0032  \\
\ m13c &120  & 50  &2.          &6.4      &0.3      &0.03       &6.4    &0.006 & 0.0027\\
\ mc14 & 120 & 80  &2.          &6.5      &1.2      &0.14       &6.1    &0.03  &0.013   \\
\ m14c &120  & 80  &2.          &6.5      &0.5      &0.06       &7.     &0.01  & 0.0067 \\ 
\ mc15 & 130 & 320 &0.1         &6.0      &1.4      &0.3        &3.0    &0.8   &0.22    \\
\ m15c &130  &320  &0.1         &6.0      &0.8      &0.2        &2.9    &0.9   &0.23    \\       
\ mc16 & 120 &100  &5.5         &6.0      &1.       &0.1        &4.5    &0.03  &0.02  \\  
\ m16c &120  &100  &5.5         &6.0      &0.7      &0.07       &4.6    &0.025 &0.019 \\
\ mc17 &120  &50   &0.17        &6.6      &1.       &0.26       &5.8    &8.e-4 &4.e-4 \\
\ m17c$^2$ &54 &200 &0.0035     &6.3      &0.15     &0.14       &3.2    &0.003 &1.6   \\
\ mc18 & 110 &100  &5.5         &6.       &1.       &0.16       &4.5    &0.044 &0.024 \\
\ m18c & 100 &120  &6.5         &6.4      &0.6      &0.1        &5.5    &0.01  &0.013 \\ 
\ mc19 &120  &100  &5.5         &6.       &1.       &0.15       &4.6    &0.04  &0.022\\ 
\ m19c &120  &100  &5.5         &6.       &0.7      &0.1        &5.0    &0.02  &0.017 \\   
\ mc20 &120  & 50  &5.          &6.3      &0.6      &0.15       &4.2    &0.016 &0.006\\
\ m20c &120  &50   &5.          &6.3      &0.6      &0.15       &4.     &0.03  &0.008\\
\ mc21 &120  & 80  & 2.         &6.5      &0.6      &0.06       &5.6    &0.033 &0.0124\\
\ m21c &120  &80   &2.          &6.       &0.5      &0.06       &5.2    &0.059 &0.016 \\   
\ mc22 &100  &200  &0.1         &4.5      &2.       &0.3        &4.     &0.95  &0.058 \\
\ m22c &100  &160  &2.          &4.       &2.       &0.09       &3.9    &9.4   &0.11\\
\ mc23 & 120 & 80  &2.          &6.5      &0.5      &0.05       &5.6    &0.033 &0.012 \\
\ m23c &120  &80   &4.          &6.       &0.6      &0.05       &5.8    &0.042 &0.015 \\
\ mc24 &130  & 320 &0.1         &6.       &1.4      &0.3        &3.1    &0.83  &0.22\\
\ m24c &130  &320  &0.1         &6.       &0.8      &0.16       &3.     &0.83  &0.22\\
\ mc25 &150  &140  &0.18        &5.       &0.14     &0.04       &6.3    &0.0094&0.01\\
\ m25c &150  &140  &0.18        &5.       &0.2      & 0.05      &6.1    &0.011 &0.01 \\
\ mc26 &120  & 50  & 1.         &6.6      &0.6      &0.06       &5.3    &0.015 &0.0042\\
\ m26c &120  & 50  &1.          &6.6      &0.6      &0.09       &4.9    &0.015 &0.0043\\   \hline    

\end{tabular}

$^1$ in \erg (calculated at the nebula); 
$^2$ calculated by an infalling model.

\end{table*}

\begin{figure}
\centering
\includegraphics[width=8.0cm]{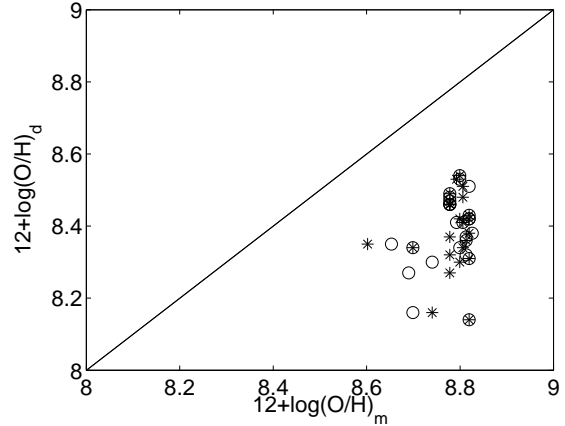}
\caption{
Comparison of log(O/H)+12  calculated by detailed modelling (m)  with
those calculated  using  indirect diagnostic calibrations by  Christensen et al (d).
open circles: model calculations; asterisks:
model calculations  referring to reddening corrected data.}
\end{figure}

\begin{figure*}
\centering
\includegraphics[width=7.4cm]{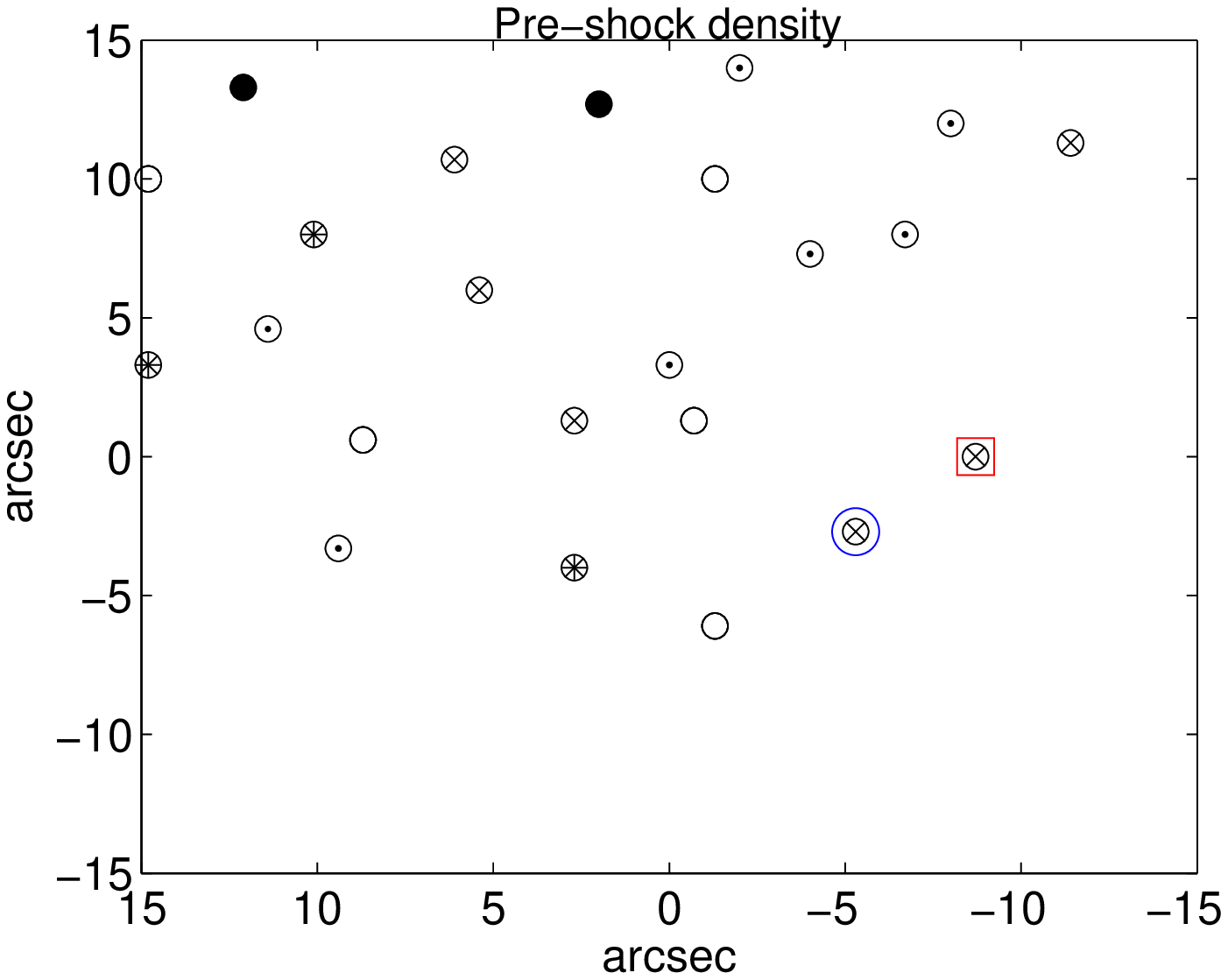}
\includegraphics[width=7.4cm]{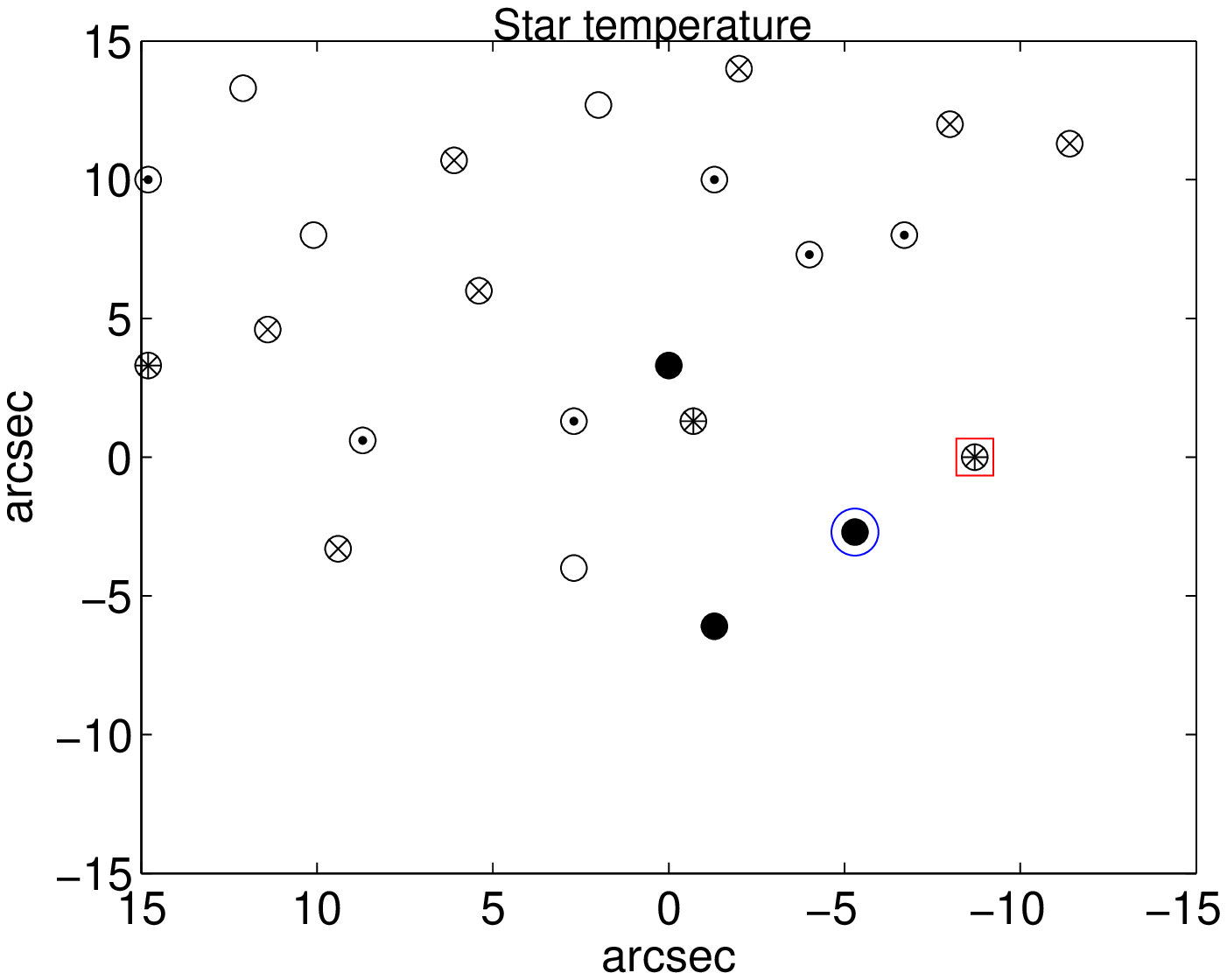}
\includegraphics[width=7.4cm]{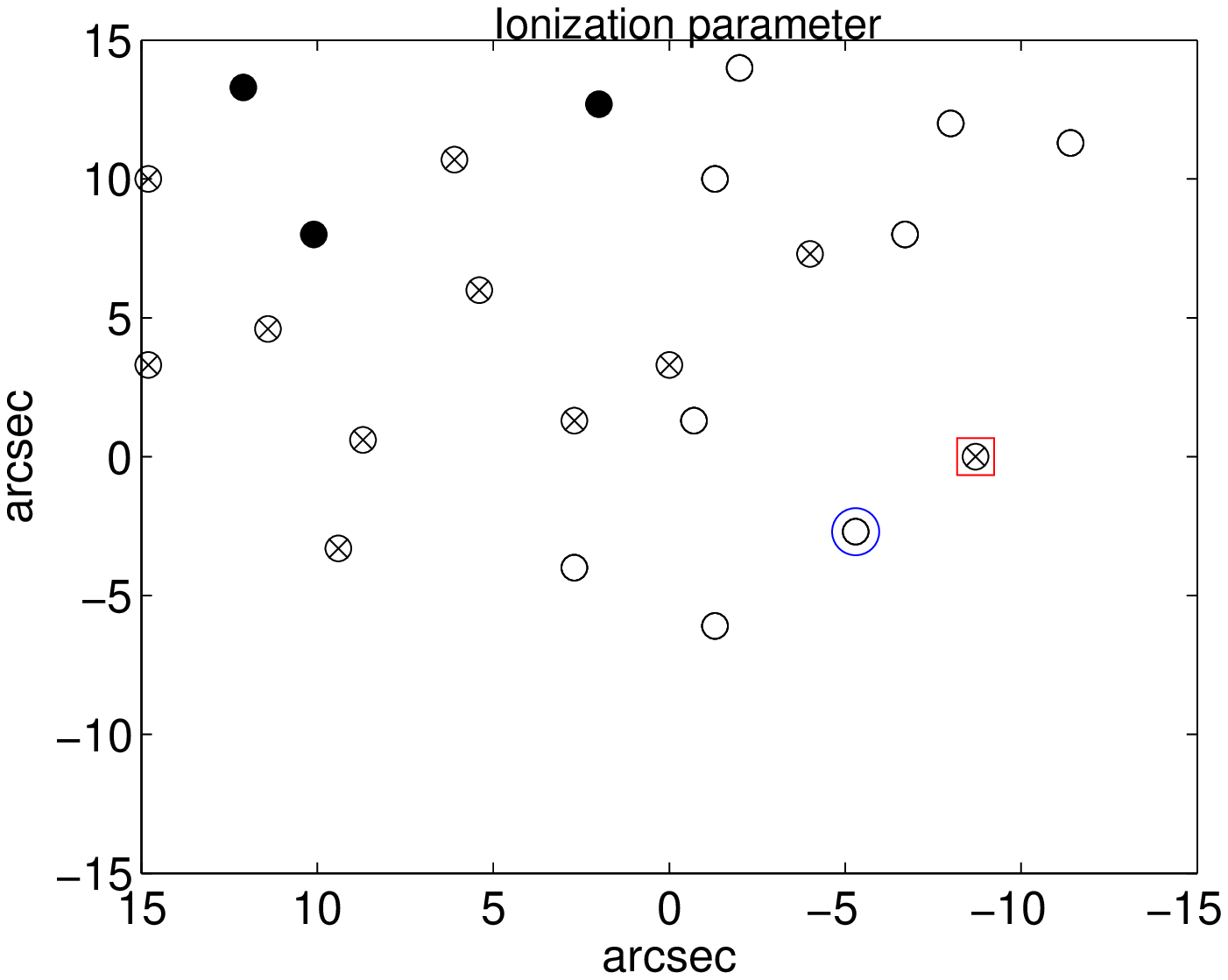}
\includegraphics[width=7.4cm]{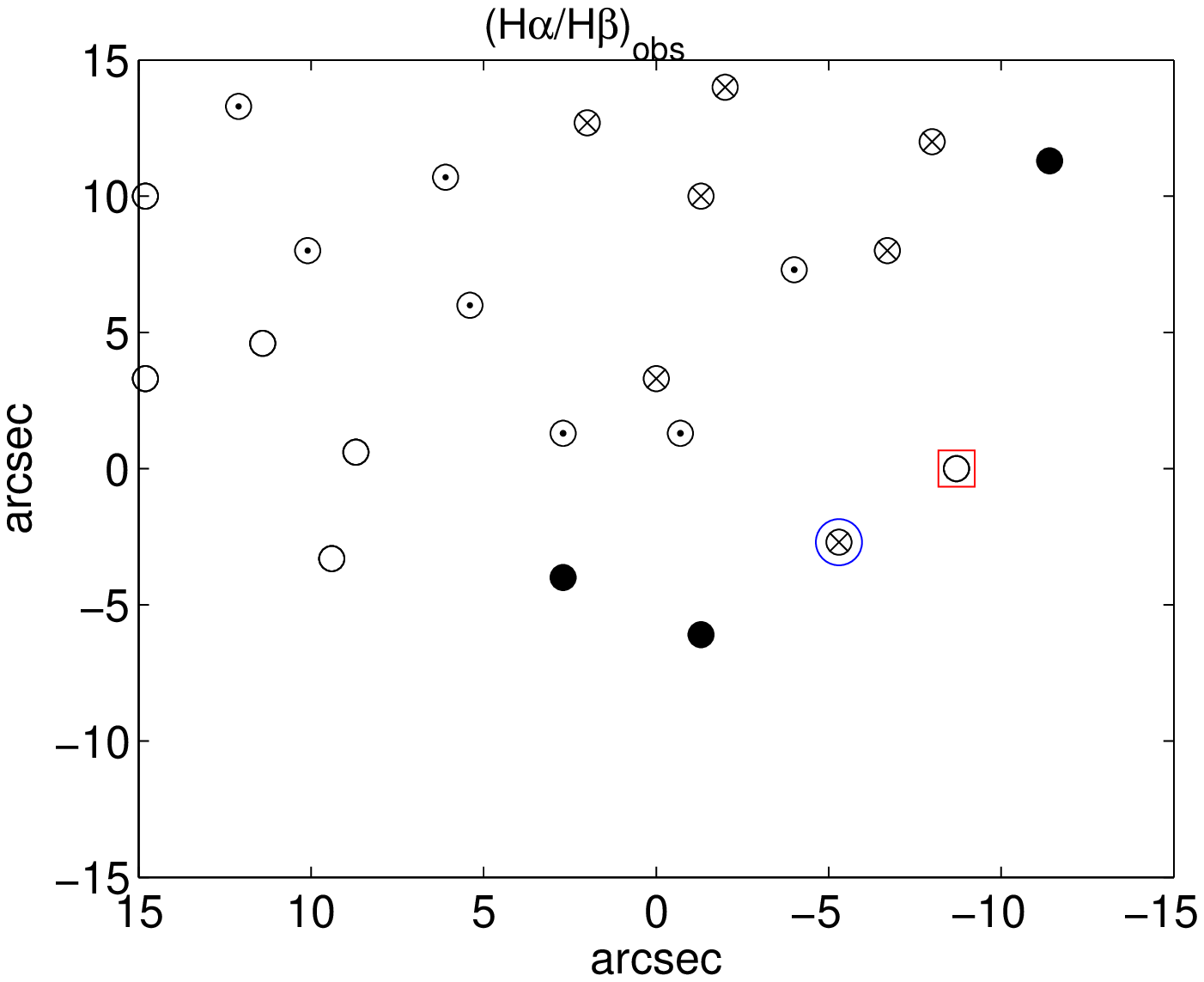}
\includegraphics[width=7.4cm]{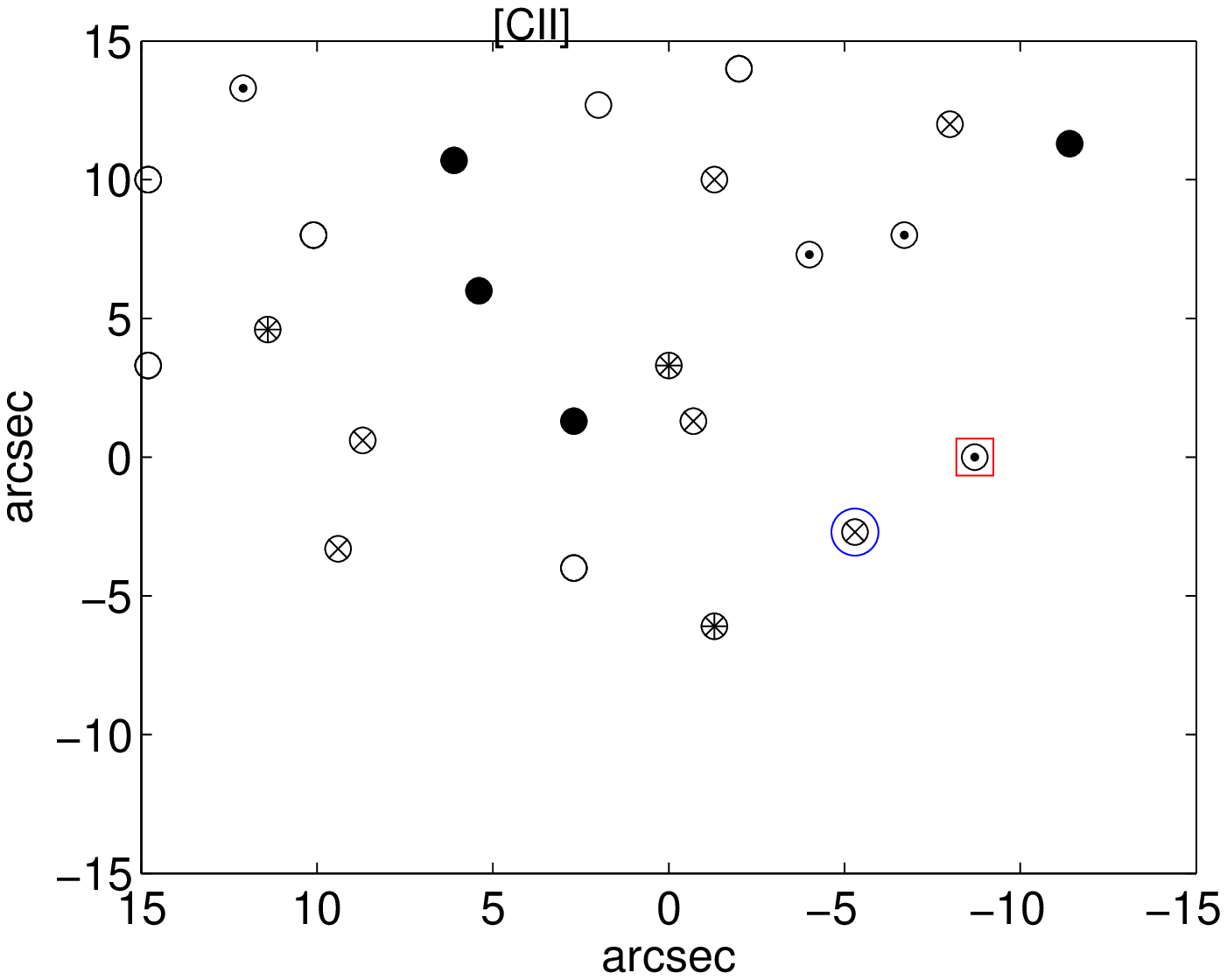}
\includegraphics[width=7.4cm]{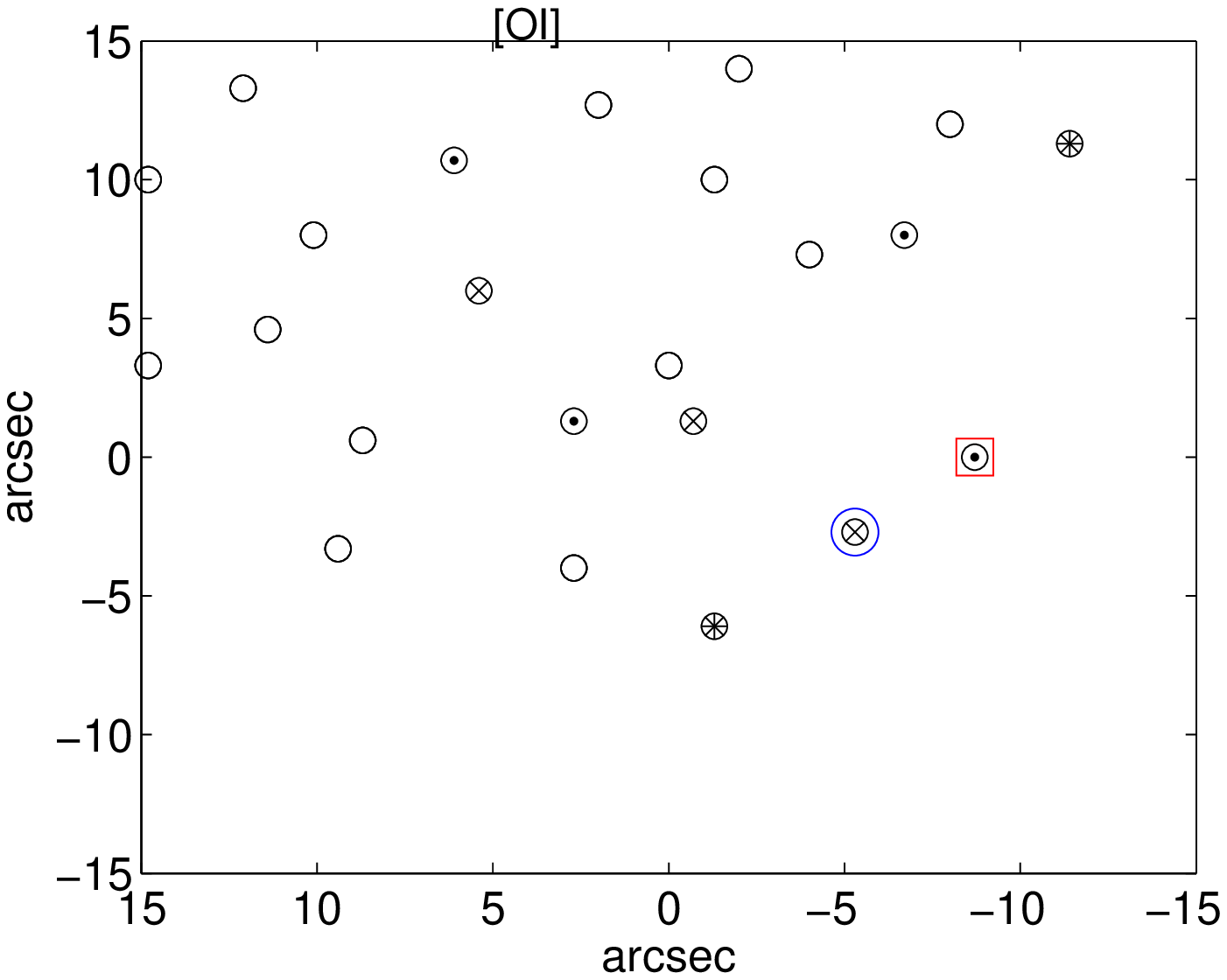}
\caption{Christensen et al data. Distribution of  the physical parameters throughout the galaxy GRB 980435
host.
\n0 (top left); 
\Ts (top right) ;
$U$ (middle left) ;
(\Ha/\Hb)$_{obs}$ (middle right);
[CII]158\mum flux  calculated at the nebula(bottom left); 
[OI]63\mum flux calculated at the nebula (bottom right);
large  blue circle : the SN place and large  red square : the region where WR stars were detected.
Symbols relative to the  different parameters are explained in Table 4.
}
\end{figure*}

\begin{table*}
\centering
\caption{Symbols in Fig. 5 diagrams}
\begin{tabular}{lcccccccccccc} \hline  \hline
\                   &left-top &right-top &left-middle&right-middle  &left-bottom          &right-bottom  \\ 
\                   & \n0     &   \Ts    &    $U$   & (\Ha/\Hb)$^1$ & [CII] line flux$^2$  & [OI] line flux$^2$ \\
\                   &\cm3     & 10$^4$ K &     -    &     -         &0.01 \erg            & 0.01 \erg   \\ \hline
\ open circle       &50       &$<$4.     &$<$0.01   &$<$3           &$\leq$0.1            & $\leq$0.1          \\
\ encircled dot     &60-80    &4.-4.9    & -        &3-5            &0.1-0.2               & 0.1-0.2         \\
\ encircled cross   &100-120  &5.-5.9    &0.01 -0.1 &5-7            &0.2-0.4              &  0.2-0.5         \\
\ encircled asterisk&140-200  &6.-6.5    &    -     &-              &0.4-0.9              &  0.5-1        \\
\ black circle      &320      &$\geq$ 7  & $>$0.1   &$>$10          & $>$0.9              &    -        \\ \hline
\end{tabular}

$^1$ observed at Earth; $^2$ calculated at the nebula

\end{table*}

 Table 3 shows that the
\Hb flux calculated at the emitting nebula is relatively high in the WR star region,
and the same should be for \Ha~  ($\sim$3 \Hb) confirming a relatively high SFR.
Metallicities in terms of O/H and N/H relative abundances are close to  solar. 
In the extreme east, in the host regions at 10.1"$\times$8.0" and 14.8"$\times$3.3" O/H  drops to
minima of 4.$\times$10$^{-4}$ and 5.$\times$10$^{-4}$, respectively.
 Fig. 4  shows that O/H values calculated by detailed modelling  exceed those obtained by Christensen et al.
using   the  strong-line method by a factor $>$2. 
  Table 3 shows that sulphur is almost underabundant everywhere ((S/H)$_{\odot }$= 2.$\times$10$^{-5}$).
 S is easily trapped into dust grains and subtracted from the gaseous
phase by factors $<$10.

Following Christensen et al we present our results in the maps of Fig. 5.
We refer to pre-shock densities, star temperatures, ionization parameters and 
observed \Ha/\Hb.
 We have chosen the symbols  showing darker areas with increasing parameter in each diagram. 
They are  described in Table 4.
Fig. 5 shows  the following features.
\Ts increases towards the south-west zone
and reaches  \Ts=8.4 10$^4$K in the SN region, slightly lower
 than  \Ts$>$ 10$^5$ K  which was calculated by Contini (2016) modelling
the Han et al (2010) survey.  Han et al  suggest that
 WR and O stars are present
in some LGRB (e.g. 980703, 990712) and  in other host galaxies (Contini 2016, table 10).
The ionization parameter has an opposite trend.  A  diluted $U$ in the SN region
indicates that
the radiation  source is far from the  emitting gas or that the photoionizing flux is prevented 
from reaching the gas by  obstructing matter.
The \Ha/\Hb observed line ratio is $<$3 in the WR zone indicating that
it is not affected by dust absorption. On the other hand, in the SN region the
line fluxes  require a reddening correction.
The preshock densities are relatively low throughout the whole galaxy,
except in the north-east zone where \n0$>$300 \cm3 are revealed.
The densities  inside the gaseous clouds     increase by a factor $>$4
due to compression downstream.

 We have found that the  line spectrum observed from  the host region  2.7" $\times$ -4.0"
in GRB 980425 is reproduced  adopting
 an  accretion model,
in agreement with Michalowski et al (2016) who claim that accretion is more adapted than outflow in starburst galaxies.
Michalowski et al recently published [CII] 158\mum and [OI] 63\mum line fluxes in the  FIR.
They claim that [CII] emission exhibits a normal radial profile while [OI] is concentrated close to the WR
zone. We have calculated the [CII] and [OI] line fluxes consistently with the  optical lines
using the same models (Table 3) in each region. Our results (Table 5, Fig. 5) show that [OI] are  weak throughout  
the host with  higher values in the regions within a band roughly oriented from north-east to south-west. 
For [CII] the radial structure can be roughly recognized.
In the WR and SN regions, the calculated values are within the mean,
suggesting that clouds  different  from those emitting the optical lines  
contribute to the FIR emission lines. According  to the results presented for the LGRB 031203 
host galaxy in Sect. 5., they could be represented by  shock dominated ($U$=0) filaments, but the data are not enough 
to constrain the models.

\begin{table*}
\centering
\caption{Flux (in 0.01 \erg) of  [CII] and [OI] lines calculated  at the nebula by the m4c-m26c models (Table 3)}
\begin{tabular}{lcccccccccccc} \hline  \hline
\  model & [CII]158\mum &[OI]63\mum &model  & [CII]158\mum &[OI]63\mum   \\ \hline
\ m4c&  0.11   &   0.207 & m16c& 1.80   &   0.098\\
\ m5c&  0.29   &   0.430 & m17c& 0.002  &   0.000\\
\ m6c&  1.03   &   0.972 & m18c& 3.29   &   0.384  \\
\ m7c&  0.21   &   0.015 & m19c& 1.34   &   0.099\\
\ m8c&  0.161  &   0.161 & m20c& 0.438  &   0.034\\
\ m9c&  0.120  &   0.009 & m21c& 0.360  &   0.033\\
\ m10c& 0.074  &   0.009 & m22c& 0.09   &   0.018\\
\ m11c& 0.588  &   0.588 & m23c& 0.72   &   0.07  \\
\ m12c& 0.403  &   0.030 & m24c& 0.176  &   0.0220\\
\ m13c& 0.269  &   0.235 & m25c& 0.028  &   0.0074\\
\ m14c& 0.871  &   0.062 & m26c& 0.105  &   0.0075\\
\ m15c& 0.176  &   0.022 &   -  &  -    &   -      \\ \hline
\end{tabular}
\end{table*}

\subsection{Hammer et al (2006)  multi-line spectra}

To check  whether the models calculated to reproduce the  line ratios  observed by Christensen et al  from
the SN and WR regions are   able to explain the data reported by other observers,  we  apply the detailed modelling 
method to the  spectroscopic VLT/FORS2 observations presented by 
Hammer et al (2006, table 1) which   contain a relatively large number 
of lines  covering an  extended range of frequencies
and of elements.
In Table 6 we compare model results (mSN, mWR and mrg4) with the data. 
The line ratios were corrected on the basis of previous considerations. 
However, the spectrum from region "4" is characterised by a relatively low \Hb flux,  leading to \Ha/\Hb=8.217.
 The observed  [OII]/\Hb =11.65  skips to 30 by reddening correction, which looks very high.
We reproduce it by $\sim$40\% adopting a very low $U$ (0.0004). The  [OIII] 4363/\Hb is overpredicted by $>$50\%
  by model mrg4.
This suggests that the abnormally high \Ha/\Hb is not only due to dust absorption, but 
 some self absorption across high density gas  has reduced the  \Hb flux   reported by Hammer et al in their table 1. 
 [ArIII]7136/\Hb observed from region "4" is well reproduced by model mrg4, while  
 [ArIII]7136/\Hb data from the WR and SN regions are underpredicted by the models by a factor of $\sim$3,
adopting Ar/H=3.3$\times$10$^{-6}$.

\begin{table*}
\centering
\caption{Modelling the   line spectra  from  Hammer et al (2006, table 2)}
\begin{tabular}{lcccccccccccc} \hline  \hline
\              & SN &SN (corr)&mSN  & WR     & WR (corr) &mWR & reg 4& reg 4 (corr)&mrg4 \\ \hline
\ [OII] 3727   &4.44  &6.23   &6.7  & 1.24   & 2.08      &3.  &11.65 &  30.13      &17.4 \\
\ [NeIII] 3869 &0.61  &0.81   &0.68 & 0.32   & 0.49      &0.49  &0.239&  0.53      &0.66\\
\ [OIII] 4363 &$<$0.033&0.039 &0.04 & 0.044  & 0.055     &0.02  &0.13 & 0.20       &0.58 \\
\ [NIII] 4640  &0.00  &0.00   & -   & 0.008  & 0.01      & - &0.00 &  0.00         &- \\
\ HeII 4686   &$<$0.055&0.056 & 0.05&0.015   & 0.015     &0.02  &0.00 &  0.00      &0.01\\
\ [ArIV] 4711  &0.00 &0.000   &0.0035& 0.01  & 0.009     &0.001  &0.00 &  0.00     &0.05\\
\ [ArIV] 4740  &0.00 &0.000   &0.0026&0.004  & 0.004     &0.001  &0.00 &  0.00     &0.04\\
\ \Hb   4861   &1.   &1.      &1.    &1.     & 1.        & 1.    &1.   &1.         &1. \\
\ [OIII] 5007+ &3.26 &3.13    &3.06 & 7.06   & 6.63      &5.26  &7.26 &  6.5       &7.27\\
\ [NII] 5755  &0.00 &0.000    &0.011 &0.005  & 0.004     &0.003   &0.00 &  0.00     &0.05\\
\ [OI] 6300+   &0.44 &0.33    &0.47 & 0.048  & 0.03      &0.16  &0.869&  0.37      &0.084\\
\ [SIII] 6312   &0.04 &0.03  &0.002 & 0.025  & 0.016     &0.001  &0.00 &  0.00     &0.053\\
\ [NII] 6548   &0.42 &0.3    &0.36 & 0.118  & 0.069     &0.1    &1.478&  0.55      &0.43\\
\ \Ha~  6563   &4.28 &3.      &3.   & 5.19   & 3.        &2.95  &8.217&  3.        &3.\\
\ [NII] 6584   &0.98 &0.68    &0.72 & 0.32   & 0.18      &0.26  &2.26 &  0.8       &0.84\\
\ [SII] 6716   &0.66 &0.44    &0.54 & 0.31   & 0.16      &0.27  &2.35 &  0.74      &0.79 \\
\ [SII] 6731   &0.93 &0.61    &0.57 & 0.24   & 0.128     &0.3   &2.56 &  0.79      &0.66\\
\ [ArIII]7136  &0.21 &0.14    &0.04 & 0.22   & 0.114     &0.03  &0.41 &  0.12      & 0.14\\
\ [OII] 7325   &0.21 &0.13    &0.2  & 0.009  & 0.041     &0.09  &0.548&  0.13      &1.\\
\ [ArIII] 7751 &0.0  &0.02    &-    & 0.006  & 0.026     & -    &0.00 &  0.00      &-\\ 
\ \Vs (\kms)   &  -  &  -     &120  &-       &    -       &150  & -    &  -        &140      \\
\ \n0 (\cm3)   &  -  &  -     &100  & -      &    -       &100  & -    &  -        &40\\
\ $D$(10$^{18}cm$)& - &  -     &0.3  &  -     &    -       &0.3  & -    &  -        &0.02\\
\ \Ts (10$^4$K)& -   &  -     &8.   &   -    &    -       &6.5  & -    &  -        &4.6\\
\ $U$          &  -  &  -     &0.004&   -    &    -       &0.032&-     &  -        &4e-4\\
\ He/H         &  -  &  -     &0.1  &   -    &    -       &0.1  &  -   &  -        &0.1\\
\ N/H (10$^{-4}$)&-  &  -     &0.3  &   -    &    -       &0.2  &  -   &  -        &0.5\\
\ O/H (10$^{-4}$)&-  &  -     &6.6  &   -    &    -       &6.6  &  -   &  -        &7.0 \\
\ Ne/H (10$^{-4}$)&-  &  -     &1. &        &     -      &1.   &   -   &  -        &0.7\\
\ S/H (10$^{-4}$)&-  &  -     &0.04 &        &    -       &0.06  & -    & -        &0.15\\
\ Ar/H(10$^{-4}$)&-  &  -     &0.033&        &    -       &0.033 &-     & -        &0.033 \\
\ \Hb (\erg) at the nebula          & -   &   -    &   0.0037&  -     &   -        &0.015&    - &  -        & 1.e-4\\ \hline       
\end{tabular}
\end{table*}

\section{LGRB 031203 optical and mid-IR lines}

The lines in the optical range from  GRB 031203 host galaxy  were observed and 
modelled in different ways in the last years (Niino et al and references therein,
Contini 2016, etc). We have reproduced the line ratios by the radiation dominated (RD) mN3 model
(Table 1),
 characterized by relatively low O/H and N/H and a relatively high \Ts.
The model accounts consistently also for the shock.

  Recently, Watson et al (2016) 
presented the first mid-IR spectrum of a GRB host (HG031203) by low and high resolution
spectroscopy with {\it Spitzer-IRS}.
 Watson et al found that the
 IRS spectra show strong high ionization fine structure emission lines such
as [SIV] 10.51\mum    which suggest a hard  radiation field in the galaxy and therefore
strong ongoing star formation and a very young  stellar population.
They claim that the  absence of PAH supports this idea as well as the  hot dust peak temperature.
We remind that high ionization lines are  strong in presence of the shocks which heat the gas
 to temperatures $\propto$ (\Vs)$^2$. Moreover, PAH grains are very small ($<$ 0.01 \mum) and easily sputtered
throughout the shock front.
We refer to the observed  [SIV]10.51\mum/[SIII]18.71\mum (=1.63) and [NeIII]15.56\mum/[NeII]12.81\mum (=15.14)
line ratios.
By the code SUMA  line fluxes from far-UV
to far-IR are calculated consistently for each model. The IR line ratios calculated by model mN3
do not reproduce the observed line ratios. On the other hand, the shock dominated (SD)
model calculated  adopting the same shock input parameters as mN3 but with $U$=0,  
approximates the data by an error of $\sim$ 50\%
([SIV]/[SIII]=3.8 and [NeIII]/[NeII]= 7.).  
 In this case the geometrical thickness of the emitting
clouds  is reduced to $D$=1.17$\times$10$^{15}$ cm.

To better understand the results, the
 profiles of the electron temperature and density within the emitting clouds and of the
fractional abundances of the H, O, Ne and S ions in  different ionization stages
are shown in Fig. 6 for both the RD (top diagrams) and the SD  (bottom diagram) models.
The RD model corresponds to the case in which the gas moves outwards from the starburst,
therefore the photoionizing radiation reaches the edge (on the right of the right top diagram) of the cloud opposite 
to the shock front (on the left of the left top diagram).  
The emitting cloud corresponding to the RD model is  divided into two halves represented by the left
and right top diagrams.  The left diagram shows
the region close to the shock front and the distance from the shock front on the X-axis scale is logarithmic.
The right diagram shows the conditions
downstream far from the shock front, close to the edge reached by the photoionization flux which is
opposite to the shock front.  The distance from the
illuminated edge is given by a reverse logarithmic X-axis scale.
The two edges of the cloud are bridged by  the secondary radiation from both sides.
In the  SD  case (bottom diagram) the gas is collisionally heated by the shock
and by  secondary radiation emitted by the slabs of gas heated  by the shock.
The gas recombines in the downstream region.
Fig. 6 shows that the optical lines calculated  by the mN3 model are emitted  from RD  clouds moving outwards, while
 the corresponding mid-IR  ones  come from cloud fragments  which are not reached by the primary radiation
flux.

\begin{figure*}
\centering
\includegraphics[width=8.8cm]{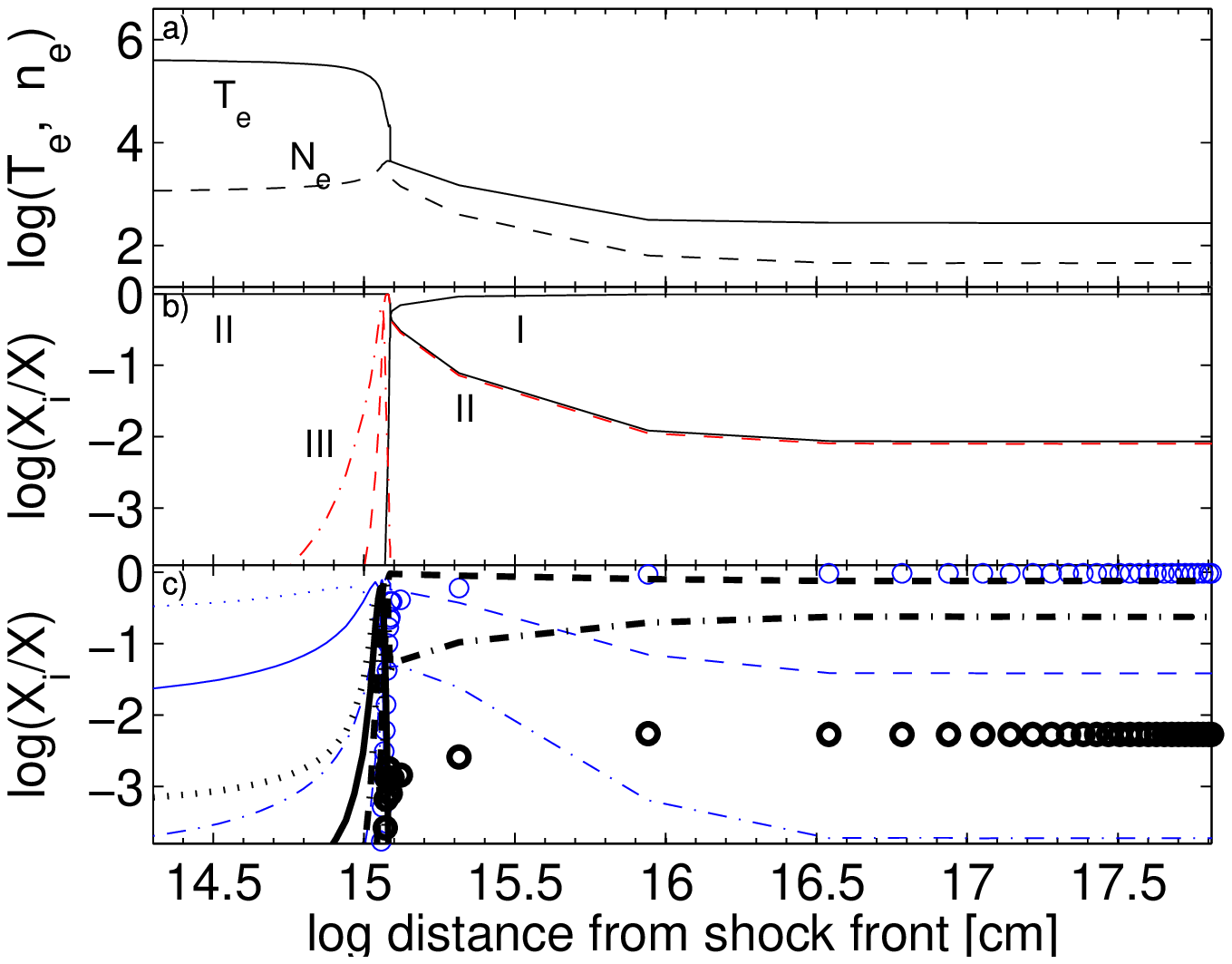}
\includegraphics[width=8.8cm]{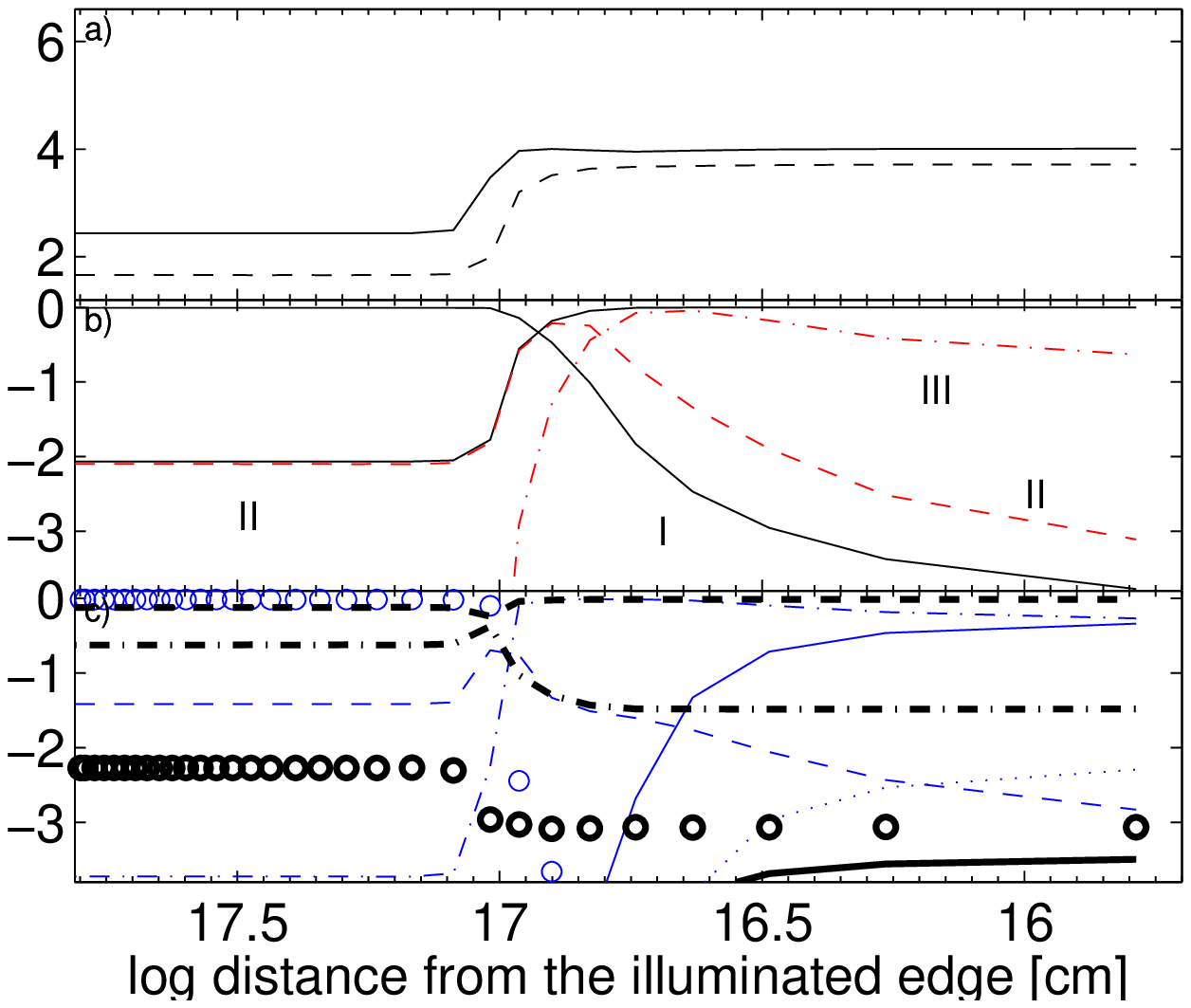}
\includegraphics[width=8.8cm]{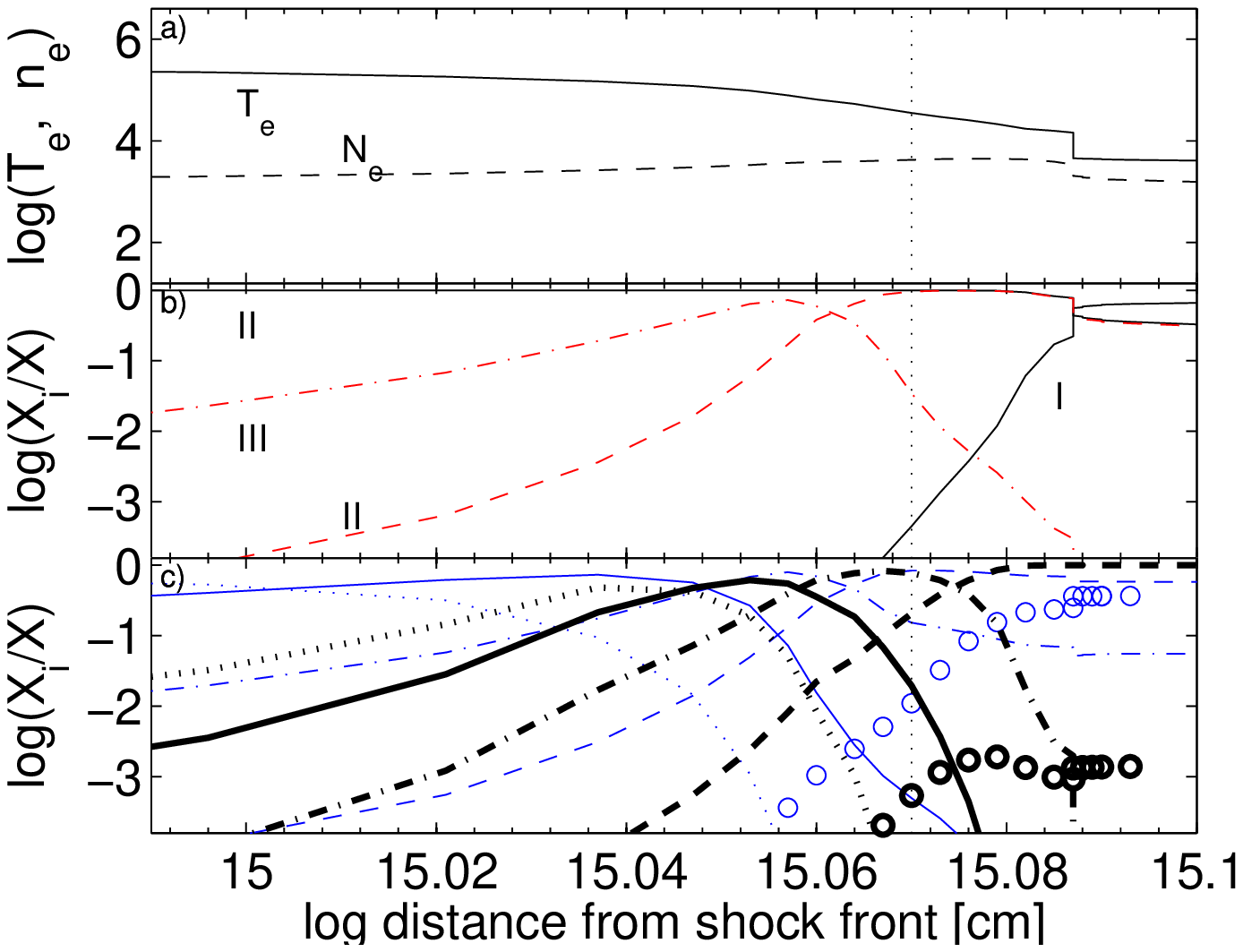}
\caption{Top  diagrams : the RD model mN3 for GRB 031203 (see text).
Top panels : the electron temperature and  the electron density throughout the emitting cloud.
Middle panels : red lines : O$^{2+}$/O (dot-dashed, III) and O$^+$/O (dashed, II): black solid lines : H$^+$/H (II)
and H$^0$/H (I).
Bottom panels : blue lines refer to Ne:
Ne$^{4+}$/Ne (dotted), Ne$^{3+}$/Ne (solid), Ne$^{2+}$/Ne (dash-dotted), Ne$^{+}$/Ne (dashed), Ne$^{0}$/Ne 
(large circles). 
Black thick lines refer to S :
S$^{4+}$/S (dotted), S$^{3+}$/S (solid), S$^{2+}$/S (dash-dotted), S$^{+}$/S (dashed), S$^{0}$/S (large circles).
Bottom  diagram : the SD model for GRB 031203. 
Symbols as in the top diagrams.
The vertical black dotted line shows the  edge of the emitting cloud}
\end{figure*}

\section{Concluding remarks}

We revisited the line spectra reported by Niino et al  for galaxies at z $\leq$ 0.41 
in order to  complete by the detailed modelling of the line ratios  the calculation of
the physical conditions and relative abundances in LGRB host galaxies in the redshift  interval 
 0.0085$\leq$z$<$0.5. 
Our results suggest that the
 high \Ha/\Hb ratios (which reach values as high as 5.5 in some of the observed spectra)  cannot  originate from  high 
density gas  which is not predicted   by the observed  [SII] doublet 
and by  strong forbidden lines in general.
Therefore,   the line ratios  were reddening corrected. 
 We have found in LGRB host galaxies at z$\leq$0.41  lower metallicities than in LGRB hosts at higher z,
suggesting that  merging is proceeding.
Detailed modelling yields  higher metallicities than those calculated by the   methods in most of the galaxies.

The investigation of different regions throughout the GRB 980425  galaxy at z=0.0085 which hosts a WR and a SN,
leads Christensen et al to claim
 that extreme values of SFR, stellar masses etc. arise only in the WR region
and that the lowest metallicity values are found  in the WR and GRB regions.  
The calculation method of metallicities
is still a debated issue and  different conclusions can be reached.
By the detailed modelling of the line ratios presented in this paper we have added more  information
 to  the  Christensen et al maps
 about  metallicities and  physical conditions in the different regions.
In particular, we have found that  the effective starburst temperature in the SN region is the highest
throughout the host galaxy.
A low $U$ reveals that in the  SN region the emitting gas
is far from the radiation source in agreement with  Fynbo et al (2000) who claim 
by high spatial resolution imaging,  that  the GRB and the associated SN 
did not occur in the regions where the  WR stars and O stars are located, corresponding to rich
and compact clusters of SFR, but several hundreds parsec away.
Moreover,  our results show   that O/H is slightly lower than solar, N/H is lower than solar and S is  depleted   
from the gaseous phase in nearly all the regions throughout the host.
Comparison of the pre-shock density map with the \Ha/\Hb one  confirms that
 high \Ha/\Hb are due to dust  reddening rather than to  self absorption  by high density gas. 
Following  Chevalier (1982) theory, two shocks are formed
after the SN explosion, one propagating towards the SN and one proceeding outward throughout the ISM.
We suggest that the line ratios observed by Christensen et al in the SN region are emitted downstream of 
the outward shock front, where  the pre-shock density and the shock velocity are suitable to the ISM.

Region  2.7"$\times$ -4.0" line spectrum (Table 2) in the GRB 980425 host, which shows an abnormally 
high [OII]/[OIII] line ratio,
could be reproduced only by an accretion model, i.e. the emitting gaseous cloud is infalling towards the
 radiation source. Accretion rather than outflow is supported by Michalowski et al in starburst galaxies.
Modelling results of [CII] 158 and [OI] 63  FIR line fluxes  observed by Michalowsky et al in the SN and WR 
regions are similar  in average  to those obtained in the other regions of the host galaxy.

The models  calculated for the SN and  WR regions on the basis of the 
Christensen et al data, were constrained only by
a few oxygen, nitrogen  and sulphur  line ratios to \Hb. We have checked them by modelling the line ratios 
observed by Hammer et al.  We have found that the same models
reproduce    satisfactorily also  the HeII/\Hb and [ArIII]/\Hb  line ratios.
High temperature stars in the SN region are confirmed. 

The modelling of  [SIV]10.51/[SIII]18.71 and [NeIII]10.6/[NeII]12.81  line ratios observed  by Watson et al
from GRB 031203 host galaxy at z=0.105 indicates that the mid-IR lines are emitted from geometrically 
thin shock dominated clouds
which are not reached by the  starburst photoionizing flux, while the optical lines are emitted from the
radiation dominated outflowing clouds.

\section*{Acknowledgements}
I am very grateful to the referee for many critical suggestions which substantially
improved the presentation of the paper.

\section*{References}

\def\ref{\par\noindent\hangindent 18pt}

\ref Allen, C.W. 1976 Astrophysical Quantities, London: Athlone (3rd edition)
\ref Anders, E., Grevesse, N. 1989, Geochimica et Cosmochimica Acta, 53, 197
\ref Asplund, M., Grevesse, N., Sauval, A.J., Scott, P. 2009, ARAA, 47, 481
\ref Blanchard, P.K. et al 2015 arXiv:1509.07866
\ref Castro-Tirado, A.J. et al 2001, A\&A,  370, 398 
\ref  Chevalier, R. A. 1982, ApJ, 259, L85
\ref Christensen, L., Vreeswijk, P.M., Sollerman, J. et al 2008, A\&A, 490, 45
\ref Contini, M. 2016, MNRAS, 460, 3232 
\ref Contini, M. 2015, MNRAS, 452, 3795
\ref Contini, M. 2014, A\&A, 564, 19
\ref Contini, M. 2003, ApJ, 339, 125
\ref Contini, M. \& Viegas, S.M. 2001a ApJS, 132, 211
\ref Contini, M. \& Viegas, S.M. 2001b ApJS, 137, 75
\ref de Ugarte Postigo, A. et al 2014, A\&A, 563, 62
\ref Drake, S. A. , Ulrich, R. K. 1980 ApJS, 42, 351
\ref Fishman, G.J., Meegan, C.A. 1995 ARA\&A, 33, 415
\ref Fynbo, J.P.U. et al 2000, ApJ, 542, L89
\ref Garnavich, P.M. et al  2003, ApJ, 582, 924 
\ref Graham, J. F., Fruchter, A. S. 2013, ApJ,  774,119
\ref Graham, J.F. et al 2015 arXiv:1511.00667v
\ref Grevesse, N. , Sauval, A.J. 1998, Space Science Reviews, 85,161
\ref Hammer, F. et al 2006, A\&A, 454, 103
\ref Han, X. H., Hammer, F., Liang, Y. C., Flores, H., Rodrigues, M., Hou, J. L., Wei, J. Y.
 2010, A\&A, 514, 24	
\ref Hjorth, J. et al. 2003, Nature, 423, 847
\ref Hook, I.M., Jorgensen, I., Allington-Smith, J.R., Davies, R.I., Metcalfe, N., Murowinski, R.G., Crampton, D. 2004, PASP, 116,425
\ref Kobulnicky, H.A., Kewley, I.J.  2004, ApJ, 617, 240
\ref Kr\"{u}hler, T. et al 2015 A\&A, 581, 125
\ref Levesque, E. M., Berger, E., Kewley, L. J., Bagley, M. M.  2010a, AJ, 139, 694	
\ref Levesque, E M., Kewley,  L.J., Berger, E., Jabran Zahid, H. 2010b,AJ, 140, 1557
\ref Michalowski, M.J. et al 2016 arXiv:1609.01742
\ref Modjaz, M. et al 2008, AJ, 135, 1136
\ref Niino, Y. et al 2016 Publ. Astron. Soc. Japan,  ArXiv:1606.01983
\ref Osterbrock, D. E. 1974 in
 Astrophysics of gaseous nebulae, San Francisco, W. H. Freeman and Co., 1974. 263 p.
\ref Paczynski, B. 1998 AIPC, 428, 783
\ref Pagel, B.E.J., Simonson, E.A., Terlevich, R.J., Edmunds, M.G. 1992, MNRAS, 255, 325
\ref Perley, D.A. et al 2016 ArXiv:1609.04016 
\ref Piranomonte, S. et al 2015, MNRAS, 452, 3293
	2005, NewA, 11, 103
\ref Prochaska, J. X. et al. 2004, ApJ, 611, 200	
\ref Rigby, J.R., Rieke, G.H. 2004 ApJ, 606, 237
\ref Savaglio, S., Glazerbrook, K., Le Borgue, D. 2009, ApJ, 691, 182
\ref Schulze, S. et al  2014, A\&A, 566A, 102S
\ref Seaton, M.J. 1975, MNRAS, 170, 475
\ref Sollerman, J., \"{O}stlin, G., Fynbo, J. P. U., Hjorth, J., Fruchter, A., Pedersen, K. 2005, NewA, 11, 103 
\ref Stanek, K.Z. et al 2003, ApJ, 591, L17
\ref Th\"{o}ne, C.C. et al 2008, ApJ, 676, 1151
\ref Th\"{o}ne, C.C. \& de Ugarte Postigo, A. 2014, GRB Coordinates Network 16079
\ref Vergani, S.D. et al 2011 A\&A 535, A127
\ref Watson, D.  et al 2014, arXiv:1010.1793
\ref Woosley, S.E. 1993, ApJ, 405, 273
\ref Xu, D. et al 2013, ApJ, 776, 98

\end{document}